PET-MRI: a review of challenges and solutions in the development of integrated multimodality imaging

**OPEN ACCESS**

**IOP** Publishing | Institute of Physics and Engineering in Medicine    Physics in Medicine & Biology

Phys. Med. Biol. **60** (2015) R115–R154    doi:10.1088/0031-9155/60/4/R115

Topical Review
# PET-MRI: a review of challenges and solutions in the development of integrated multimodality imaging

Stefaan Vandenberghe[1] and Paul K Marsden[2]

[1] Department of Electronics and Information Systems, MEDISIP, Ghent University-iMinds Medical IT-IBiTech, De Pintelaan 185 block B, B-9000 Ghent, Belgium
[2] Division of Imaging Sciences and Biomedical Engineering, School of Medicine, King's College London, St Thomas' Hospital, Westminster Bridge Road, London, SE1 7EH, UK

E-mail: Stefaan.Vandenberghe@ugent.be and Paul.Marsden@kcl.ac.uk
Received 18 July 2014, revised 3 October 2014
Accepted for publication 13 October 2014
Published 4 February 2015

**Abstract**

The integration of positron emission tomography (PET) and magnetic resonance imaging (MRI) has been an ongoing research topic for the last 20 years. This paper gives an overview of the different developments and the technical problems associated with combining PET and MRI in one system. After explaining the different detector concepts for integrating PET-MRI and minimising interference the limitations and advantages of different solutions for the detector and system are described for preclinical and clinical imaging systems. The different integrated PET-MRI systems are described in detail. Besides detector concepts and system integration the challenges and proposed solutions for attenuation correction and the potential for motion correction and resolution recovery are also discussed in this topical review.

(Some figures may appear in colour only in the online journal)

## 1. Introduction and overview

The subject of this review is a clear description of the major challenges and different steps taken towards the integration of PET and MRI technology. Both hardware and processing aspects necessary for simultaneous quantitative PET-MRI are discussed in this topical review.

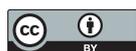 Content from this work may be used under the terms of the Creative Commons Attribution 3.0 licence. Any further distribution of this work must maintain attribution to the author(s) and the title of the work, journal citation and DOI.
0031-9155/15/04R115+40$33.00   © 2015 Institute of Physics and Engineering in Medicine   Printed in the UK    R115

*1.1. Multimodality imaging with PET-CT and PET-MRI*

The idea of combining PET and MRI imaging devices in a single system was first suggested in the early-mid 1990s (Hammer 1990, Hammer *et al* 1994). PET detectors capable of measuring in strong magnetic fields (Shao *et al* 1996) and prototype MRI-compatible PET scanners capable of imaging small animals simultaneously with MRI started to appear soon afterwards (Christensen *et al* 1995, Shao 1997). Only after a period of about 15 years of developments, have human systems capable of sequential (Zaidi *et al* 2011) or simultaneous PET and MRI acquisitions of the whole body become available commercially (Delso *et al* 2011). The main reason for this slow progress is that the integration of PET and MRI is much more complex (Cherry 2009) than the evolution from standalone CT and standalone PET systems towards multimodality PET-CT. One of the motivations for integrating PET and MRI is the huge success of PET-CT but there are clear differences. The integration of PET and CT in one system was a relatively simple process with a clear advantage for clinical studies and patient throughput (Beyer and Townsend 2006). All current commercial PET-CT scanners consist of essentially unmodified standalone PET and CT scanners mounted in-line in a common gantry and with a single patient couch. Spatially registered CT and PET scans are performed sequentially and it is assumed that the patient does not move between the two acquisitions. From the CT image the required information for attenuation correction is derived (Kinahan *et al* 1998). This information obviates the need for lengthy transmission scans and has led to a significant increase in patient throughput compared to standalone PET scanners.

For PET-MRI the technical problem of system integration is more challenging due to the presence of magnetic fields. There are different options for the system configuration (Delso and Ziegler 2009). A technically relatively simple approach for PET-MRI is to place the MRI and (minimally modified) PET components in-line in a similar configuration to PET-CT (Zaidi *et al* 2011, Kalemis *et al* 2013). This solution does not allow simultaneous acquisition and will have lengthy acquisition times, on the other hand it does not require the development of a completely new MRI compatible PET system. In PET-MRI the real goal has been to fully integrate both systems without reducing the performance of the PET and the MRI components. Full integration is the desired option as it results in acquisition times comparable to MRI acquisitions, it has a much smaller footprint and it enables novel simultaneous imaging protocols. A fully integrated system can however only be fulfilled once the PET detectors are MRI compatible and sufficiently compact to be integrated inside the main magnet. Besides this the interference problems between the MRI and PET components also need to be minimized. Before giving an overview of the different proposed solutions, the major challenges for integrating the two modalities are described.

*1.2. Challenges in the integration of PET and MRI*

*1.2.1. MRI compatible PET detectors.* The main challenge in combining PET and MRI has been the development of compact MRI-compatible PET detector technology. Detectors for the 511 keV gamma rays emitted following positron decay are the key component of all PET scanners (Lewellen *et al* 2008, Peng and Levin 2010). The vast majority of PET systems constructed to date have been based on some configuration of an inorganic scintillation crystal array optically coupled to a small number, typically four, of photomultiplier tubes (PMT). The PMTs detect the light that is emitted when a 511 keV gamma ray interacts with the scintillator and enable the position, energy and time of the interaction to be determined. Unfortunately, the performance of standard PMTs is severely degraded in even a weak magnetic field of several mT (Pichler *et al* 2006). The main challenge in developing a PET scanner that can operate within or in close proximity to an MRI scanner, where the field within the magnet bore is typically between 0.5 and 10 T, has therefore been either to develop a PET detector that can operate within such a high field, or to find other ways to circumvent this problem. Whilst many different solutions have been implemented for small animal imaging (with less need for good energy and timing resolution), the development of viable human PET-MRI systems capable of simultaneous PET and MRI acquisition has been associated with the emergence of robust and reliable MRI compatible solid state photodetectors such as avalanche photodiodes (APDs) (Renker 2007) and more recently silicon photomultipliers (SiPMs) (Britvitch *et al* 2007, Roncali and Cherry 2011). These devices have many desirable properties that make them of great interest as alternatives to photomultiplier tubes in many applications (including non-MRI-compatible PET), but as they are essentially insensitive to large magnetic fields, they were required for developing integrated simultaneous PET-MRI with high performance. It is important to note that there are important differences between small animal and clinical imaging regarding the MRI field. Most human PET-MRI scanners have been developed around 3 T MRI magnets. However for small animal systems there is a wide variety of MRI systems depending on the type of studies required, from low field 0.3 T systems up to 11 T, with 7 T being one of the most used field strengths. The lower field systems are more oriented towards anatomical imaging while the stronger fields are used for more challenging imaging tasks.

*1.2.2. Space and time constraints.* The majority of small animal and human PET-MRI research projects are aimed at a purpose-designed MRI-compatible PET scanner to be located inside the bore of the MRI magnet, allowing simultaneous PET and MRI acquisition. The choice between sequential and simultaneous configurations is linked to the uncertainties over clinical applications, but for human systems the trend appears to be towards simultaneous acquisition. In this case the limited space for the PET detectors inside the MRI bore adds another important challenge for the detector technology. Besides the potential for simultaneous PET and (f)MRI imaging there are also two practical benefits for a fully integrated clinical system compared to a sequential system (Delso and Ziegler 2009). The first factor is determined by the typical longer acquisition times in MRI compared to CT: a CT scan will last about 15 s–1 min while typically several MRI sequences are acquired in one imaging session and the total acquisition time is about 20–40 min. The most recent evolutions in PET technology (3D, Time-of-Flight and longer axial FOV) (Muehllehner and Karp 2006) have led to a strong increase in effective system sensitivity and reduced the total acquisition time for brain imaging to 3–15 min and for whole body scans within the range 10–20 min. In a sequential system the PET and MRI need to be at a significant distance from each other and therefore the total acquisition time will be the sum of the PET acquisition time and MRI acquisition time. Compared to PET-CT this leads to a very long acquisition time and patient throughput will be limited. In a simultaneous system the acquisition time will be mostly determined by the slowest component, i.e. the MRI acquisition and throughput will be comparable with current MRI scanners. A second factor is the size of the system, MRI systems are already quite large and the addition of a PET in sequential mode requires a room larger than most current imaging rooms. Both factors (acquisition time and required installation space) are less important in small animal imaging. The first commercial sequential PET-MRI systems have recently become available for small animal imaging (Nagy *et al* 2013). To date there are not yet any commercial simultaneous small animal systems.

*1.2.3. Quantitative imaging.* Besides the complexity of the integration of hardware, PET-MRI has another major technical challenge compared to PET-CT. Quantitative PET imaging requires the linear attenuation coefficients (at 511 keV) of the object to perform attenuation and scatter correction (Turkington 2000). In a PET-CT system the CT image is not only used for anatomical information but is also easily transformed to linear attenuation coefficients. In a PET-MRI system the derivation of attenuation correction maps from either MRI images, emission data or transmission imaging is complex (Martinez-Möller and Nekolla 2012, Keereman *et al* 2013). Although several methods have been developed the problem is not yet completely solved and remains an active area of research.

*1.2.4. Clinical use.* In addition to the many technical obstacles and the attenuation correction problems, another reason why it has taken so long for human PET-MRI systems to develop as quickly as PET-CT is the current absence of definitive clinical applications. Early pre-clinical prototype systems were developed with the aim of acquiring complementary functional and anatomical information concurrently to create a powerful research tool. PET provides highly specific data on molecular pathways, and MRI provides images of anatomy with excellent soft tissue contrast combined with an ever expanding range of functional and molecular measurements. Whilst there is no shortage of potential clinical research applications in humans and there are clear benefits in terms of dose reduction, the benefits of PET-MRI in routine clinical patient management are not yet so clear. The most promising areas are studies where MRI is already the preferred anatomical imaging modality (eg. in the brain, breast and abdomen) where MRI's superior soft tissue contrast make it preferable to CT), and an improved throughput and patient experience is made possible by acquiring PET and MRI images in the same scanning session. Another important limiting factor for integration in clinical routine is the large budget for acquiring a system and the high maintenance costs. Now that human systems are available it will be possible to see exactly where the utility of PET-MRI in patient management and clinical research lies (Pichler *et al* 2010, Jadvar and Colletti 2014).

In this review the technical problems associated with combining PET and MRI into a single system, and the approaches that have been used to address these, are described, along with outlines of some representative complete imaging systems (see table in Disselhorst *et al* (2014) for a complete overview of all developed systems). Most of the developments described are PET-related, reflecting the majority of work that has been carried out to date, i.e. MRI-compatible PET scanners operating within an essentially unmodified MRI scanner. Only recently with the advent of commercial human systems have developers started to modify aspects of the MRI in order to achieve close integration and minimise interactions between the PET scanner and the MRI gradient and RF coils within the magnet bore. Now it has been established that the basic compatibility issues are solvable, attention is turning to the operational and quantitative aspects of (in particular, human) PET-MRI systems. Reliable and accurate methods for attenuation correction in PET-MRI are essential and must have comparable accuracy to CT-based methods. Whilst PET data acquisition is essentially passive and limited to the acquisition of data from a single radiopharmaceutical at a time, an MRI imaging session typically involves acquisition of images acquired with many different pulse sequences resulting in many datasets all providing very different information.

## 2. Technical challenges for combining PET and MRI

Combining PET and MRI into a single device is a challenging task requiring many new developments. Before going into detail in the different technical challenges for the detectors and systems we will first describe in the next section the potential benefits of an integrated PET-MRI system.

*2.1. Motivations for combining PET and MRI*

Some of the motivations for PET-MRI are driven by the experience and limitations of PET-CT which is now the current standard platform for PET studies. Combined PET-CT has proved extremely successful because it provides improved accurate spatial registration between PET and CT images (compared to standalone systems), because of the logistical advantages of acquiring PET and CT scans in a single imaging session and also because the CT scan data can be used to correct the PET scans for the effects of photon attenuation much more rapidly (a whole body CT takes about 2–3 min) than previous methods using radionuclide sources (Beyer and Townsend 2006). Current PET-CT systems are however far from perfect (Kinahan *et al* 2003). A single PET image typically takes several minutes to acquire and so is blurred due to respiratory and other sorts of patient movement (Liu *et al* 2009). A CT scan on the other hand is essentially a snapshot. Mismatches between the blurred PET and snapshot CT result in significant artefacts in attenuation corrected PET images, particularly around the region of the diaphragm. There are many potential solutions to the problems of patient motion and PET-CT mismatch. These often make use of dynamic, cine or gated CT data, however, the radiation dose associated with spiral CT means that it is unlikely ever to be as flexible as MRI in acquiring rapid dynamic whole body anatomical data over long durations (although new x-ray technology and reconstruction methods do promise significant reductions in CT radiation dose). PET-CT has particular limitations for small animal imaging (Ford *et al* 2003) where the soft tissue contrast is very poor compared to that obtained in humans, and for repeated studies the radiation dose to the animal can become very large. In studies investigating the effect of novel anti-tumour drugs the dose may be large enough to effectively provide radiotherapy treatment thus obscuring the response of the tumour to the drug. In many applications, for example staging of lung cancer (one of the main applications of PET-CT), CT is the anatomical imaging modality of choice (MRI performs poorly in the lungs) and PET-CT is therefore the ideal combination. There are applications however, notably the brain, breast, prostate and abdomen, where the additional information delivered by CT is limited.

Based on the limitations of current clinical and preclinical PET-CT systems the different motivations for developing a combined PET-MRI system are :

(a) MRI is more flexible than CT due its plurality of sequences and can for different body regions provide improved contrast
(b) There are ways to use co-registered (both spatially and temporally) MRI data to improve the quality of PET data.
(c) The total delivered dose can be reduced significantly as no dose is required for anatomical imaging.

Besides the clinical potential there is also a large research potential for PET-MRI:

(a) The potential for new complex functional studies. MRI is not limited to anatomical imaging, but has also functional imaging options like BOLD-fMRI, DWI, DCE, spectroscopy. Furthermore one can administer hyperpolarised compounds and other exogenous agents.
(b) Investigate interventions and dynamic processes in the body simultaneously using PET and MRI.
(c) Kinetic and dynamic studies using information (like blood flow) from other modality.
(d) New specific PET-MRI multi modality tracers.

From this list it is clear that a completely functional PET-MRI system involves more than the development of MRI compatible detectors and hardware system integration (Gaertner *et al* 2013). Now that the first whole body PET-MRI systems have become available, the acquisition process needs to be optimised. The development of co-registered MRI and PET data in time and space will require significant modifications of the PET and MRI acquisition, reconstruction and visualisation. A next step is the optimisation of disease-specific clinical protocols. This involves the selection of the best combination of sequences and PET-tracer and determination of imaging time.

*2.2. Interference between PET and MRII*

Compared to the standard PET technology significant modifications are required at the detector and system level before integration is possible. We first explain the design and performance of current PET and MRI systems, then describe in detail interference problems and the potential options for modifying the PET and MRI systems.

*2.2.1. State-of-the-art PET scanners.* Currently available human PET systems for clinical and research imaging all follow a very similar overall configuration comprising an annulus of scintillation detectors (as shown in figure 1) that surround the patient in order to detect pairs of 511 keV gammas in coincidence (Muehllehner and Karp 2006). The inner diameter of the scintillator ring is typically in the range of 85–90 cm, in the axial direction the detectors extend 15–25 cm and with a patient aperture of typically 70–80 cm diameter. The scintillation detectors that make up the ring are known as 'block detectors' each comprising a segmented block of inorganic scintillator coupled to an array of, usually four, photomultiplier tubes (PMTs). Processing of the signals from the PMTs allows the position, energy and time of a gamma ray interaction in scintillator to be determined. The scintillator material most commonly used is *Lutetium Oxyorthosilicate* (LSO or LYSO) which is chosen because of its high effective Z and density, coupled with very good light output and timing properties, resulting in high spatial resolution, sensitivity and temporal resolution. Detectors have a thickness in the range of 1.5–3 cm to have sufficient stopping power at 511 kev. The most recent systems now have detectors with good energy resolution (11–12 percent) for limiting the detection of scattered coincidences originating in the patient. There is also the trend towards excellent time resolution. This allows the accurate measurement of the time difference between the arrival of both photons. With this information the position of the annihilation can be localised along the line joining the two detection positions. The uncertainty in position is proportional to the uncertainty in measuring the difference in gamma arrival times. A typical 'time-of flight' resolution of 500 ps results in a positional uncertainty of 7.5 cm, and this additional information is used to improve the quality of the reconstructed image (Karp *et al* 2008). Depending on the length of the object a typical PET acquisition requires one or more (step and shoot acquisition) bed positions. The majority of studies are for oncological investigations and will acquire an image of the whole torso. The most recent scanners have high sensitivity enabling the acquisition of a whole body in under 20 min. For brain imaging one bed position will cover the whole object, which also enables dynamic imaging with time frames below one min.

　　The majority of small animal PET scanners also have a cylindrical geometry, but there is a larger variety in the dimensions. The systems typically have a diameter in the range of 10–20 cm enabling the acquisition of mice and rats. The axial length typically has a range from 3–12 cm. Compared to human systems, thinner scintillators are used (0.5–1.5 cm). Excellent energy and timing resolution are less important for small animal imaging as the objects are small, resulting in limited scatter and no benefit of time-of-flight PET.

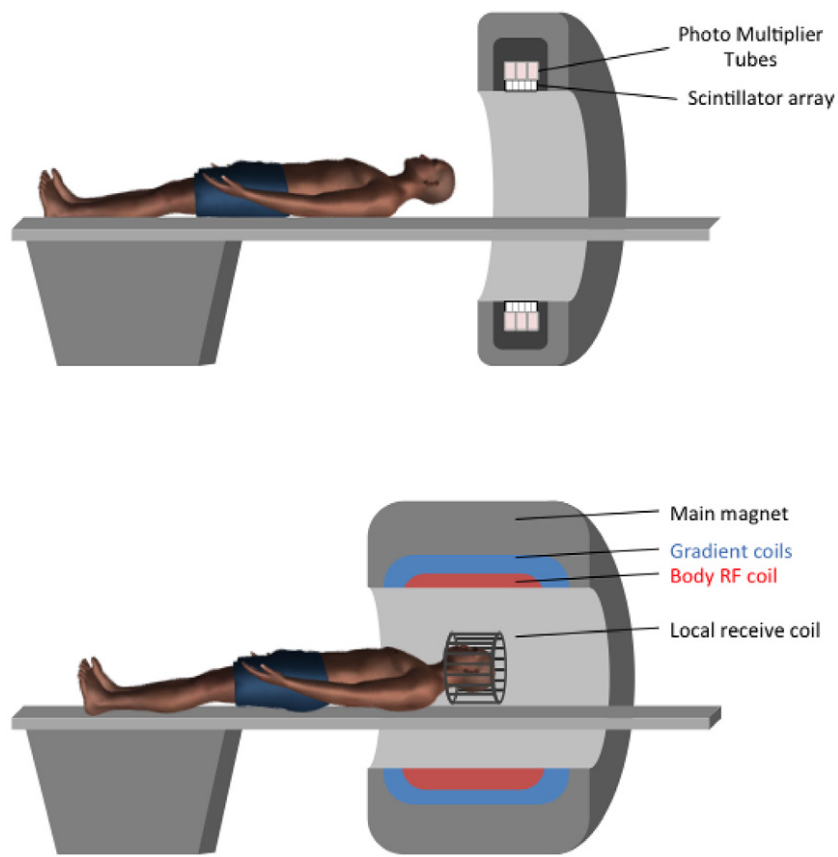

**Figure 1.** The geometry, components and structure of a standard PET scanner (top) and of a standard MRI scanner (bottom).

*2.2.2. State-of-the-art MRI scanners.* A whole body MRI scanner (Carpenter and Williams 1999, Blamire 2008) has the following key components: the main magnet, the gradient coils and RF coils for sending and receiving the signal. The magnet will have a field strength ranging from to 0.2 T–7 T. Nearly all current clinical systems use a superconducting magnet of 1.5 T–3 T. The most important criterion for the magnet is the need to maintain a good homogeneity in the static field $B_0$ of typically below 4 ppm over a spherical region of 50 cm diameter which effectively defines the field of view. The image SNR nominally increases linearly with $B_0$, with higher field strengths also providing improved spectroscopic resolution. The diameter of the magnet bore is usually 60–70 cm, and in general good homogeneity becomes harder to achieve as the diameter of the magnet increases. The $x$, $y$, $z$ gradient coils are mounted concentrically within the magnet bore. The gradient coils allow the strength of the $B_0$ field to be varied linearly as a function of $x$, $y$ and $z$, so providing spatial localisation of the MRI signal and determining the image spatial resolution and geometrical accuracy. Gradient performance is a key factor for fast dynamic imaging. This is characterised by the maximum gradient amplitude they can achieve (field gradient) as well as by the rate of change of the gradient (slew rate). The high power dissipated by high performance gradients requires air- or water cooling.

Installed concentrically within the gradient set is the whole body transmit coil (as shown in figure 1) that provides uniform RF excitation at the Larmor frequency over the whole field view. Whilst large volume transmit coils can also be used to receive MRI signals, the SNR of the detected signal is considerably increased by the use of small application-specific local receive coils (surface coils) that can be placed very close to the volume of interest (eg spine, neck, knee, wrist, shoulder, breast etc). Receive coils can also be used in the form of phased arrays, that allow imaging of large volumes whilst retaining the high SNR.

Most small animal MRI scanners have a similar setup (Carpenter and Williams 1999, Blamire 2008). The inner bore of the magnet typically ranges from 16–30 cm. Inside this bore the gradient coil set is inserted which reduces the inner bore to a range of 9–20 cm. Inside this volume RF coils are positioned for imaging specific parts of the animal. Typical field strengths are 4.7 T, 7 T, 9.4 T. The most advanced systems are based on fields of 11.7 and 15.2 T. Recently new small animal systems with lower field strengths (range of 0.5 T–3 T) have become available. These systems do not require cryogen cooling and have smaller stray fields, eliminating the need for dedicated MRI rooms with quench pipes.

Whereas most PET imaging sessions acquire essentially a single image, with minimal scope for optimising acquisition parameters, a clinical MRI session usually involves many examinations with different contrast mechanisms (sequences) performed sequentially to examine different aspects of the disease. The parameters of each investigation can be adjusted to optimise the complex trade-offs between spatial resolution, SNR, field of view, speed and contrast, and imaging sessions are usually longer than for PET with 30–45 min being typical.

*2.2.3. Physical integration of systems.* The combination of MRI and PET scanners is highly challenging (Delso and Ziegler 2009). First of all there is a challenging physical constraint. Both whole body PET and MRI systems have an inner bore diameter of 60–80 cm and an outer bore above one meter. The integration of both modalities in one system capable of simultaneously imaging the same object therefore requires the reduction in size of one of the modalities. The most obvious choice is the reduction of the thickness of the current PET detectors. Once the detector is compact, it can be integrated inside a big bore MRI (inner diameter above 70 cm) without reducing the FOV below the size required for whole body imaging. For small animal systems similar limitations of physical constraint are present (Judenhofer *et al* 2008): inner clear bore MRI diameters range from 9–30 cm. Although the small animal MRI systems with the largest diameter are a factor 2–3 × more expensive than the ones with a small bore, system integration efforts for small animal imaging were focussed on the large bore systems due to the significant thickness of PET detectors.

*2.2.4. Ways in Which MRI Can affect PET.* There are three main effects (see table 1) that prevent the technology currently used in standalone PET systems being able to function within or near to an operating MRI scanner. The high static magnetic field, quickly changing gradient fields, and radiofrequency signals from the MRI scanner prevent the normal operation of photomultiplier tubes and may induce interference in the front-end electronics of PET detectors.

(a) Main magnetic field
    Photomultiplier tubes such as those used in the standard PET block detector design will not function in even very weak static magnetic fields. The magnetic field perturbs the paths of electrons moving from the photocathode down the dynode chain to the anode resulting in a loss of gain. Attempts can be made to shield PMTs using steel or mu-metal but such measures area only effective against relatively weak fields. There are designs for field-insensitive PMTs and some position-sensitive PMT (PSPMT) configurations are

**Table 1.** Interference effects of MRI on PET performance.

| MRI | Effect on PET | Solutions | Consequences |
| --- | --- | --- | --- |
| $B_0$ | Changes path of electrons in readout | Replace PMTs by APDs, SiPMs | Higher cost<br>More channels<br>Reduced timing (APD) |
| $B_1$ | Heating, vibration | Redesign of electronics (no conductive components)<br>Temperature control | Additional complexity |
| RF | Interference with electronics | RF shielding around PET | Increased eddy currents and heating |

less sensitive to magnetics fields, but none of them will tolerate the fields of several Tesla encountered in an MRI system. The solution to this problem has therefore been either to remove PMTs from the high magnetic field region or to replace PMTs with new light detectors that have better tolerance to magnetic fields.

(b) Gradient fields

Gradient fields are switched rapidly at a frequencies of the order 1 kHz and so, due to the greater skin depth at lower frequencies, are more difficult to shield than the higher frequency RF (120 MHz @ 3 T). These rapidly switching magnetic fields can induce eddy current loops in any conductive components introduced into the magnet bore, including PET circuitry. In addition to signal interference, these can lead to heating and mechanical vibration. Because of the very confined space in the magnet bore, temperature sensitive solid state photodetectors detectors and electronics are likely to be placed very close to the gradient set, and mechanical vibration places much greater emphasis on the robustness of all aspects of detectors and electronics than would normally be necessary. The solution is to redesign the PET readout system and minimise the vibration effects.

(c) RF interference

Any electronics situated within the magnet bore may be susceptible to RF interference generated by the MRI transmit coil. This effect is responsible for the drop in PET count rate that is observed in many MRI-compatible PET systems during MRI acquisition. RF shielding is, however, more effective at this higher frequency range and so in principle PET detectors and electronics within the magnet bore can be enclosed in a conducting shield to reduce RF interference.

### 2.3. Ways in Which PET Can affect MRI

Conversely the introduction of PET detectors inside the gradient coil and magnet can lead to interference. (see table 2)

(a) Susceptibility artefacts

Any small differences in magnetic susceptibility (Schenck 1996), caused by PET scanner components within the magnet bore may result in an inhomogeneity in the main magnetic $B_0$ field. It may also affect the linearity of the gradient fields beyond the level that can be accurately corrected by shimming . Particular items are scintillation crystals, any RF shielding materials, or dense gamma shielding materials such as lead or tungsten (Strul *et al* 2003), and in particular electronic components containing ferromagnetic materials (Wehrl *et al* 2011). Gradients may induce eddy currents in for example shielding materials and electronics which in turn can distort the $B_0$ and gradient linearity. To the extent possible, only nonmagnetic materials should be used in the PET detectors to maintain the homogeneity of the magnetic field and to minimise the creation of susceptibility artifacts.

Table 2. Interference effects of PET on MRI.

| PET component | Effect on MRI | Solutions | Consequences |
| --- | --- | --- | --- |
| Scintillators | $B_0$ non-uniformities | Use of MRI compatible PET scintillators | |
| Gamma shielding | Eddy currents lead to distortion and non-linearity | Alternative gamma shielding materials | Higher cost |
| PET electronics and power cables | Interference with RF detection | RF shielding around PET | |

(b) RF interference

The NMRI signals generated in the body in response to excitation by the MRI $B_1$ field are extremely weak, requiring the MRI receive coils to be of very high sensitivity and the complete MRI scanning room to be Faraday shielded. Any unwanted sources of RF occuring during the coil receive period, that are in the MRI detection frequency range (e.g. 120 MHz at 3 T) will degrade and distort the received signal and so degrade the final MRI images. Many of the frequencies present in modern digital electronics, eg clock pulses, are indeed in this frequency range so great care must be taken to minimise interference by shielding of the PET components. The same RF shielding works in both directions. The shielding material needs to be carefully chosen and the material itself may result in inhomogeneities and eddy currents. Unshielded power cables may carry noise currents into the MRI coil circuitry.

*2.4. Effective integration of both modalities*

To design a simultaneous PET-MRI system based on the currently available MRI systems the key component is a compact MRI compatible PET detector. This detector should allow the construction of a PET system with the following properties: the system should have comparable PET performance compared to current PET-CT systems, it can be inserted inside the MRI scanner and interference between PET and MRI is minimal. An overview of the developments in the field of MRI compatible PET detector technology is given in the following section.

## 3. MRI compatible PET detectors

The most efficient detector for 511 keV gamma rays is an inorganic scintillator coupled via a light guide to a photodetector, and nearly all MRI-compatible PET detectors have been based on some modification of this standard configuration. The first requirement for MRI-compatibility is therefore that the scintillator material itself does not lead to susceptibility artefacts. Fortunately, the most widely used PET scintillators LSO, LYSO and BGO, have a magnetic susceptibility similar to that of human tissue (Yamamoto *et al* 2003) and have been demonstrated to have negligible effects on MRI. Scintillators containing gadolinium, such as GSO and LGSO, are not suitable, however these are less commonly used for PET nowadays.

As mentioned above, photo-multiplier tubes (PMT) will not operate inside the magnetic field of an MRI. Early developments in MRI-compatible PET were therefore focused on positioning the PMTs outside the magnetic field, using long optical fibres (typically up to 1 mm in diameter, i.e. Much thicker than those used in communications) to transfer light from the scintillator to PMTs situated in a low magnetic field region (one of the first developments of such a detector is shown in figure 2). The use of light guides has long been a standard technique for reading out scintillators in high magnetic fields for particle physics experiments. In addition to the complexity associated with handling large numbers of optical fibres, these systems suffered from poor energy and temporal resolution because there are different factors affecting the scintillation light. Depending on the incident angle a fraction of the light isotropically emitted from the crystal bottom will not be transmitted, there is also attenuation along the length of the finer and dispersion causes transit time variation, resulting in slower rise time and poor timing resolution. MRI-compatible PET detector development is therefore now focussed on replacing the PMTs with a new generation of solid-state photodetectors which can result in only a minimal sacrifice in PET or MRI performance.

The compact format of these solid state devices compared to PMTs means that they can be used to construct annular PET scanners that will fit within the bore of an essentially unmodified MRI system. In order to achieve, this the PET diameter is in general less than that of a standard PET configuration. In this situation, the ability of the detector to resolve the gamma interaction position in the radial direction ('depth-of-interaction' capability) is expected to be of value, but this feature has not been implemented on any complete systems to date. In all other ways, notably efficiency, spatial, energy and temporal resolution, the required performance parameters for an MRI-compatible detector are essentially the same as for any other PET detector.

Placing the photodetector within the magnet means that any front-end electronics required to be close to the detector must also be placed there. Great care must be taken with electronic circuit layouts, component choice and RF shielding in order to avoid the various types of interference between PET and MRI outlined above. Based on this there are three major detector concepts for inserting a PET system inside an MRI which are illustrated in figure 3. Siemens has designed a slight variation on the most direct APD based readout by inserting analog electronics preamplifier) close to the APD.

Some preliminary investigations of the performance of Si, HPGe and CdTe solid state detectors in high magnetic fields and in MRI scanners have been made (Burdette *et al* 2006, Harkness *et al* 2011, Cai and Meng 2012), however the potential of these detectors for PET imaging is not yet clear.

### 3.1. Solid-state photodetectors

In a solid-state photodetector (SSPD) high field regions are created within the structure of a semiconductor device when an external voltage is applied. The two most relevant types of SSPD for PET-MRI applications are those with internal gain, notably the avalanche photodiode (APD) and silicon photomultiplier (SiPM) also know as a Geiger mode APD (G-APD) or multi pixel photon counter (MPPC). Because electron-hole pairs created by incident optical or UV photons are actively collected by an electric field within the volume of the device, these devices can have a much higher quantum efficiency than PMT photocathodes and, because the path traversed by charge carriers is very short, they are immune to magnetic fields of the strength encountered in MRI. Pichler *et al* (2006) and Pichler *et al* (1997) demonstrated that an APD suffers no change in gain when placed in a 9.4 T magnetic field, independently of orientation, and similar results have been obtained for SiPMs at 7 T (Espa na *et al* 2010). SSPDs are very compact and can be manufactured in small arrays with individual detector elements in the range of $1 \times 1$ mm$^2$ up to $5 \times 5$ mm$^2$ providing many options for reading out arrays of small scintillation crystals including one to one crystal-detector coupling and light sharing schemes where the light from a large number of crystal elements is read out by a smaller number photodetectors elements.One of the main trade-offs in detector design is that one-to-one crystal-photodetector coupling avoids the detrimental effects of sharing light between several photosensor elements, however it can lead to a very large number of electronic readout channels with consequent high power and heat dissipation requirements (not to mention cost) which can be problematic in the confined space of the MRI bore. Maintaining SSPDs at a stable low temperature is also important to keep dark count rates (thermally generated signals) at a manageable level.

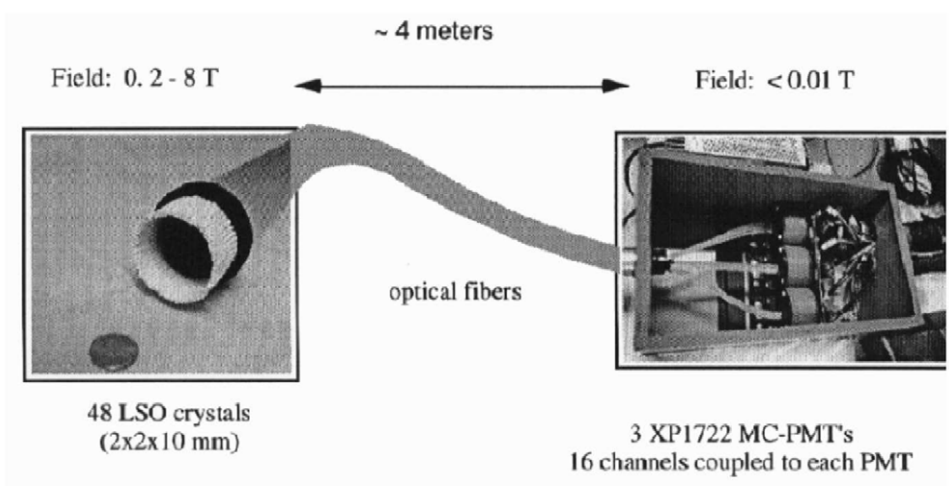

**Figure 2.** Long optical fibers are used to guide the scintillation light outside the MRI system, which enables the use of photomultiplier tubes for PET (reproduced with permission from Shao (1997)).

*3.1.1. APD based detectors.* In an avalanche photodiode the operating voltage is set below the device breakdown voltage so the charge signal obtained is proportional to the original number of electron–hole pairs produced and to the energy deposited in the scintillator by the incident gamma. The gain increases with the applied voltage, however at very high gains there is a deterioration in SNR and the optimum operating range corresponds to a gain of only about 50–150, with voltages in the range of 100–1000 V. APDs can have a quantum efficiency (QE) as high as 80 percent (at 420 nm—the peak emission wavelength for LSO) however an APD/LSO combination can typically achieve a temporal resolution only of a few ns (compared with <600 ps FWHM for a PMT/LSO combination) which is inadequate for time-of-flight PET. A major challenge of APDs is that the gain is very sensitive to changes in temperature ($\sim$ >3%/deg C) and voltage ($\sim$ >10% $V^{-1}$), so these must be very carefully controlled. To read out the very small signals produced, low noise readout electronics is required and this must be placed as close to the detector as possible. Short optical fibres have also been used in combination with position sensitive APDs (just outside the FOV) to position the electronics at a distant position. An example of an APD detector with short optical fibres in shown in figure 5 and one with direct coupling of APD to scintillator in figure 6.

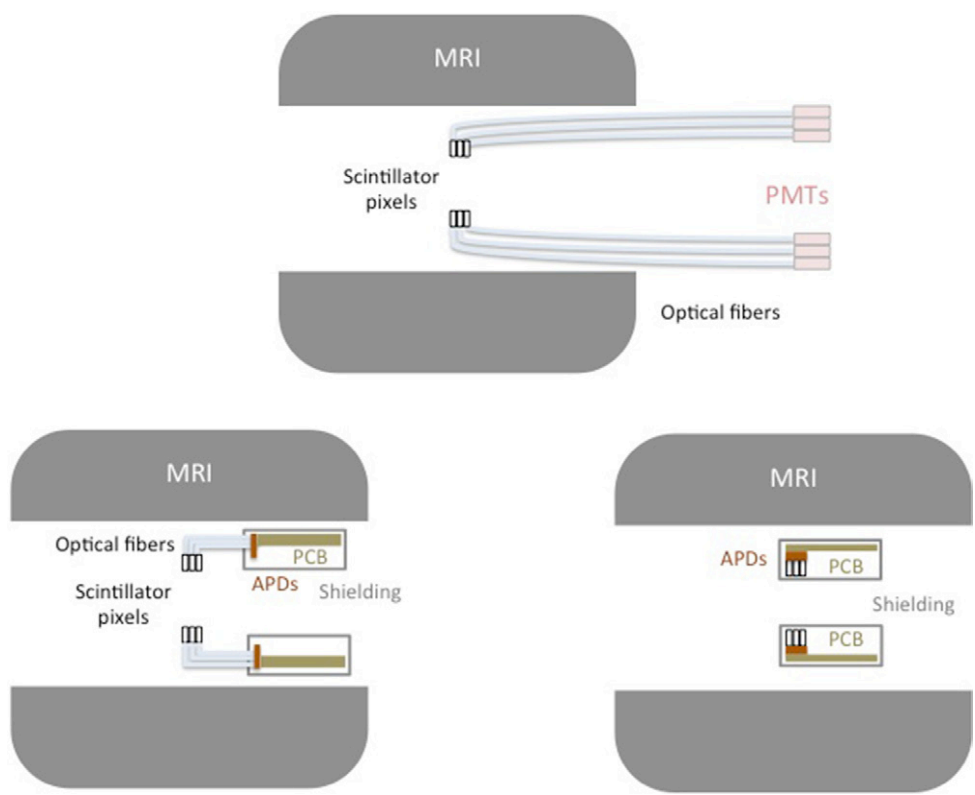

**Figure 3.** Three detector concepts for integration of PET and MRI: use of long optical fibers used to couple scintillators to PMTs outside the magnet (top), APDs directly attached to scintillator elements (bottom left), scintillators attached via short optical fibers to APDs (bottom right) (adapted with permission from Catana *et al* (2006)).

For these reasons APDs have not achieved widespread use in conventional PET systems, nevertheless, as there is more experience in the use of these devices than for SiPMs, they are used in several current generation small animal MRI-compatible PET scanners and in the simultaneous whole body PET-MRI scanner, the Siemens mMR, which is described below.

*3.1.2. SiPM based detectors.* If the operating voltage is increased beyond a threshold level, then the creation of charge carriers in the detector results in a Geiger discharge. This signal is very large but is no longer proportional to the energy deposited by the incident gamma. An SiPM (Roncali and Cherry 2011) consists of APDs operated in Geiger mode, however the proportionality with energy is regained by dividing the active area of the detector into an array of very small GAPD elements or 'cells' each of dimensions typically $20 \times 20$–$60 \times 60$ um$^2$. Each cell acts independently so, provided the number of incident optical/UV photons is much less than the total number of cells (so that the probability of one cell to be hit by two photons will be small), then the summed output from all the cells is proportional to the energy deposited in the scintillator. Good energy linearity is important as photodetector outputs need to be compared in crystal identification schemes.

SiPMs exhibit gains comparable to that of a PMT for very modest bias voltages of 30–100 V, so the need for very low noise front end electronics is removed. The high SNR and speed of the

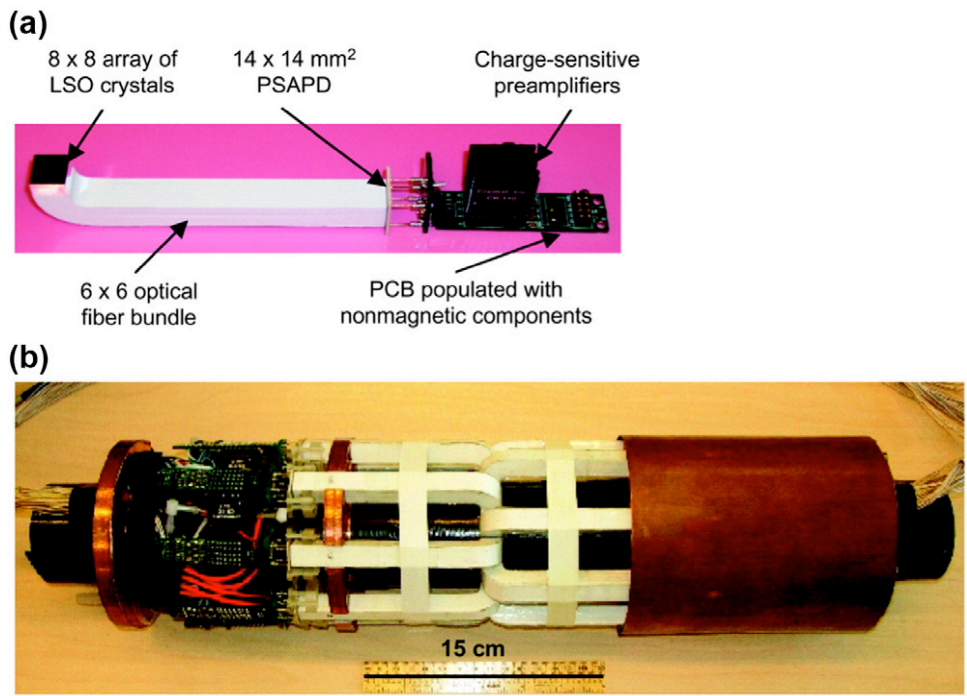

**Figure 4.** The light from LSO crystals is guided to APD light sensor (top view (*a*)), different arrays are combined into a complete small animal system (bottom view (*b*)) (reproduced with permission from Catana *et al* (2006)).

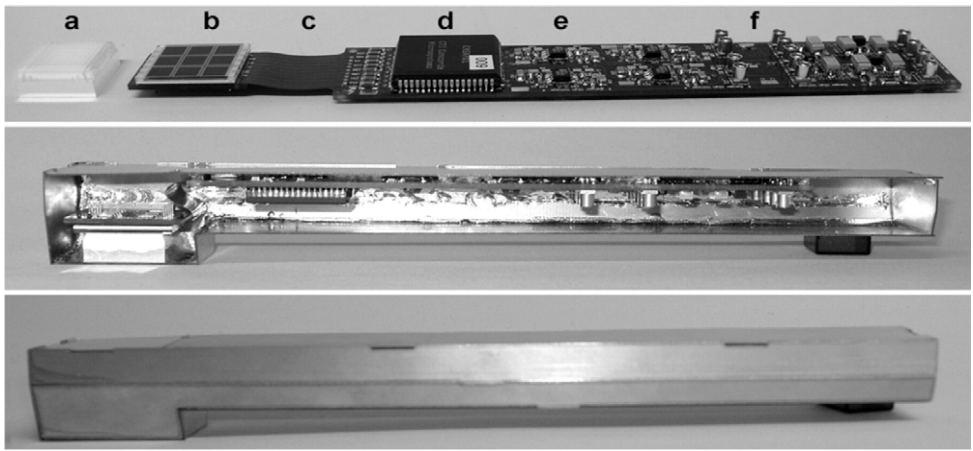

**Figure 5.** The upper figure shows the crystal block and light guide (*a*) coupled to an APD (*b*) connected to preamplifier (buffers and connectors (*d*)–(*f*). The middle picture shows the detector mounted into its housing and at the bottom the shielded PET module is shown (reproduced with permission from Judenhofer *et al* (2007)).

Geiger breakdown result in excellent timing properties with a temporal resolution of ∼100 ps having being reported for room temperature measurements on single LaBr3:Ce(5%) crystals individually coupled to SiPMs (Schaart *et al* 2010). Because of a thinner depletion layer, and because the sensitive area is <100 percent, the photon detection efficiency (PDE, which is more relevant than the QE for SiPMs) is usually lower than can be obtained with APDs, i.e. typically <40% with LSO.

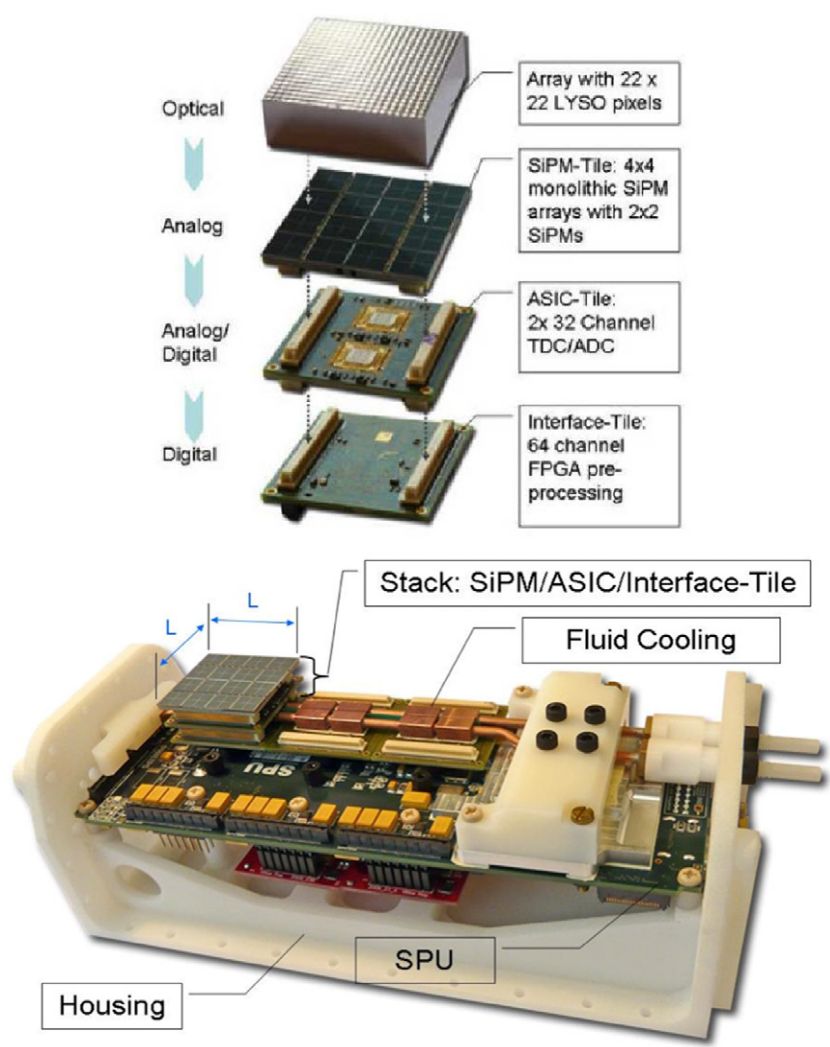

**Figure 6.** ASIC readout of analog SiPMs and module based on analogue SiPM (reproduced with permission from Schulz *et al* (2009)).

SiPMs can be fabricated into arrays of increasing dimensions and with very high gain, excellent temporal resolution and low bias voltage and promise to address the various shortcomings of APDs. These properties of course make SiPMs very attractive not only for PET-MRI but also for conventional PET, though as for APDs there is a strong dependance of gain on temperature and especially bias voltage in addition to various trade-offs required to maintain acceptable levels of dark counts. A recent development is the digital SIPM (Degenhardt *et al* 2009) where the signals from individual microcells are summed digitally (i.e. a binary '1' for every cell that discharges) using electronics integrated on the sensor chip. This results in improved performance (though the PDE may be lower) and lower temperature sensitivity. The sensor itself performs some of the functions (e.g. time to digital conversion and trigger logic) that would otherwise have to be performed by an additional readout ASIC (application specific integrated circuit). An example of an analog SiPM detector and module in shown in figure 7 and one based on digital silicon photomultipliers in figure 8.

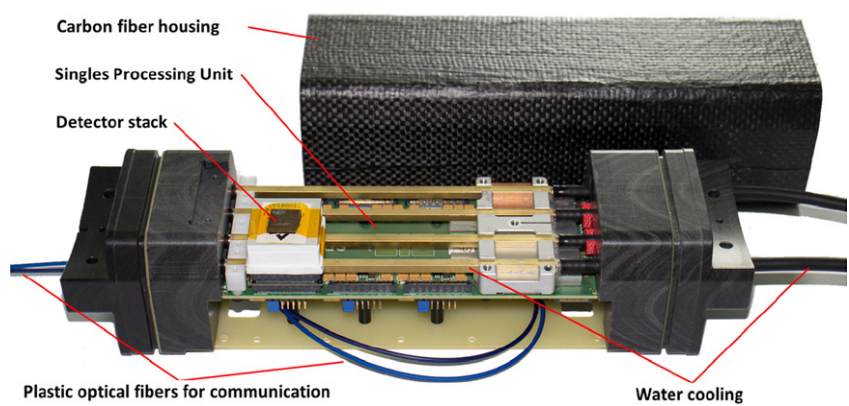

**Figure 7.** An MRI compatible PET detector module based on digital SiPM (reproduced with permission from Wehner *et al* (2014)).

## 4. MRI compatible PET systems

### 4.1. Small animal systems based on PMTs

*4.1.1. Early systems using light guides and optical fibres.* Experiments using scintillators, light guides and PMTs to demonstrate both the reduction in positron range caused by a strong magnetic field and the feasibility of a simultaneous PET-MRI scanner were reported in 1995 by Christensen *et al* (1995). Similar experiments had been presented by the same group previously using photodiodes (Hammer *et al* 1994), however it was thought that, due to the lack of any conducting or magnetic components in the field of view, the lightguide approach would be preferable for the construction of a combined PET-MRI scanner such as had been previously proposed by Hammer (1990). Around the same time, Buchanan *et al* (1996) presented a (non-imaging) scintillator/light guide combination that was able to perform $^{18}$F-FDG uptake measurements simultaneously with $^{31}$P spectroscopy of the Langendorf perfused rat heart in a 9.4 T spectroscopy magnet. Scintillation light from a NaI crystal, situated adjacent to the rat heart within the bore of the magnet, was transferred via a long light guide to a magnetic field insensitive PMT situated in a field of ∼50 mT just outside the magnet. The device enabled changes in the PET and MRIS measurements, resulting from interventions such as hypoxia or ischaemia, to be directly correlated with each other in the same experiment. The first version of the microPET small animal PET-only scanner (Cherry *et al* 1997) utilised short 15 cm optical fibres to transfer scintillation light from LSO crystal arrays to position sensitive photomultiplier tubes, and it was soon realised that by extending the length of the optical fibres, the microPET detector configuration could easily be adapted to work in a magnetic field. Several prototype systems were constructed with long (∼3 m) 2 mm diameter optical fibres linking a single detector ring (5–12 cm in diameter) of LSO crystals to 3 or more PS-PMTs in a low field region (<10 mT) outside the magnet (Shao 1997, Mackewn *et al* 2010). The total absence of magnetic or conducting components in the field of view resulted in excellent MRI compatibility in a range of MRI scanners (0.2 T–9.4 T), however scintillation light was attenuated by the optical fibres (typically the amount of light was reduced by a factor 5–10) and combined with the single ring configuration this resulted in poor timing resolution, energy resolution and sensitivity.

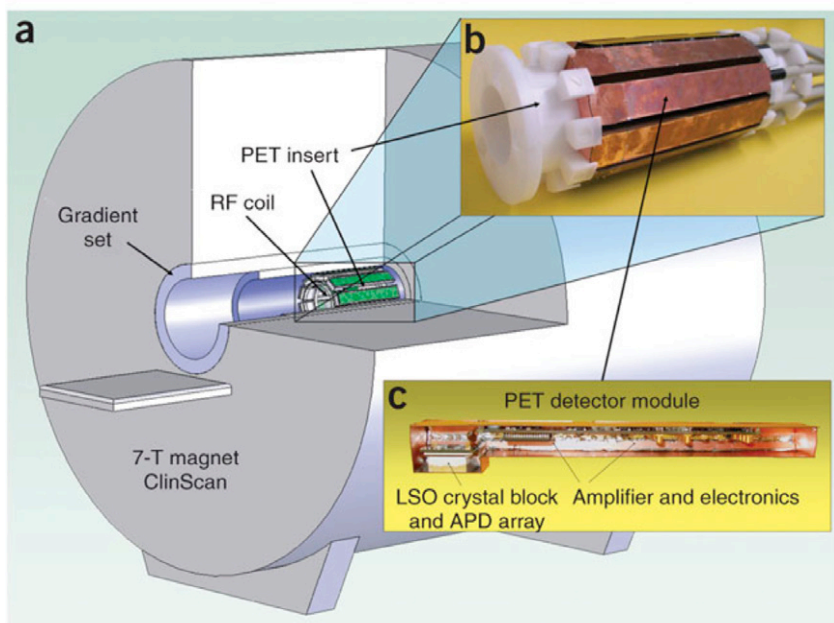

**Figure 8.** PET insert inside a small animal MRI scanner (*a*), the MRI-compatible PET insert with detector modules (*b*) and PET detector module based on LSO scintillators, APDs and preamplifier inside copper shielding (*c*) (reproduced with permission from Judenhofer *et al* (2008)).

The optical fibre approach was further developed by Raylman *et al* (2006) who used optical fibre bundles with a 90 degrees bend in order to accomodate an axially extended PET field of view within a conventional MRI scanner, however more recent approaches using PMT/fibre combinations have also considered other magnet configurations. Yamamoto *et al* (2010) have constructed an integrated PMT/fibre system where again an axially extended PET field-of-view is obtained by using a 90 degree angled light guide to channel light from the crystal arrays into 75 cm optical fibres. The PET annulus is located in the field of view of a 0.3 T permanent magnet with the fibres passing out of the magnet through a hole in the magnet yoke enabling the PMTs to be located in a 0.3 mT magnetically shielded region immdiately behind the yoke. The slanted light guides still result in ~90% light loss, and the coarse crystal segmention results in a transaxial PET spatial resolution of 2.9 mm FWHM, however despite the limited capablities of the 0.3 T MRI, this very simple and practical configuration allows excellent access to the animal and negligible PET-MRI interference.

*4.1.2. Sequential small animal systems.* A straightforward way to capitalise on well characterised existing PET and MRI technology is to adopt a sequential PET-MRI configuration. The Mediso nanoScan® PET-MRI imaging system (Nagy *et al* 2013) comprises a small animal PET scanner (LYSO arrays read out via relatively field insensitive 256-channel PS-PMTs) mounted in line with a compact 1 T permanent magnet MRI and with a common animal holder. Whilst the relatively low field MRI does not have all the capabilities of more highly specified MRI systems, a straightforward replacement of sequential PET-CT with sequential PET-MRI has many advantages for small animal imaging, including dramatically improved soft tissue contrast and reduced radiation dose.

*4.1.3. Split-magnet and field cycling small animal systems.* Another approach is to modify the MRI system itself in such a way that PMTs only have to operate in an acceptably low field. In the Cambridge split-magnet system (Hawkes *et al* 2008, Buonincontri *et al* 2013), a 1 T superconducting magnet is split in the axial direction leaving an 80 mm gap into which a modified microPET Focus 120 small animal PET scanner (Laforest *et al* 2007) can be inserted. Optical fibre bundles transfer scintillation light in the radial direction from the scintillator arrays in the magnet bore to PMTs situated in a low field region just outside the magnet—to achieve this, the fibre bundles are extended from 10 cm in the standalone PET scanner design to 110 cm long. In order to create a sufficiently low field for the PMTs the magnet is self-shielded either side of the gap, and mild steel shields are also inserted. With further mu-metal shielding, the PMTs experience a field of just 1 mT and overall there is only a small reduction in PET performance compared with the standalone system. As it is not desirable to place attenuating objects (other than a relatively low-attenuation RF coil) between the object being imaged and the PET detectors, the gradient set is also split (Poole *et al* 2009). Split magnet configurations like this have previously been used for interventional human MRI systems (Schenck *et al* 1995), however the configuration does inevitably compromise the gradient performance and the homogeneity of the main magnetic field to a small extent.

Field-cycled MRI systems use combinations of resistive electromagnets with the capability of varying field strengths dynamically. They can produce images of comparable quality to clinical superconducting systems, and allow new types of MRI contrast. Because the magnets can be rapidly switched on and off (i.e. in just 30 ms) combined PET-MRI operation is possible by performing PET and MRI alternately with very little interference between the modalities. In the prototype system described by Bindseil *et al* (2011), standard PET block detectors are placed in a gap in the 0.3 T polarizing magnet in a similar way to the Cambridge split magnet arrangement, however only minimal mu-metal shielding of the detectors is implemented and no optical fibres are needed to distance the detectors from the high magnetic field (in this system the gradient set is not split). Interleaved PET and MRI acquisitions have been demonstrated with a repetition time (TR) for a combined PET/MRI sequence of 2752 ms with PET acquisition occuring for ∼1/3 of the time.

The split magnet and field-cycling approaches continue to be developed, however they have not been widely adopted, and remain the only examples of the use of non-standard magnet configurations for PET-MRI.

*4.2. Small animal systems based on solid state photodetectors*

Little development is now happening with PMT-based systems. When these early systems were first developed (largely by PET investigators) the sources of interference were not well understood, APDs and SiPMs were not widely available and, although multiplexing methods have been demonstrated e.g. Graziose *et al* (2005) the complexity of reading out large numbers of crystals elements resulted in limited axial extent and sensitivity. As will be seen below it appears that the various PET-MRI interference issues are surmountable with detector components situated within the MRI and so it is not necessary to sacrifice PET or MRI performance with significantly modified PET or MRI designs. For this reason nearly all recent developments of simultaneous systems have been focussed on APD and SiPM approaches.

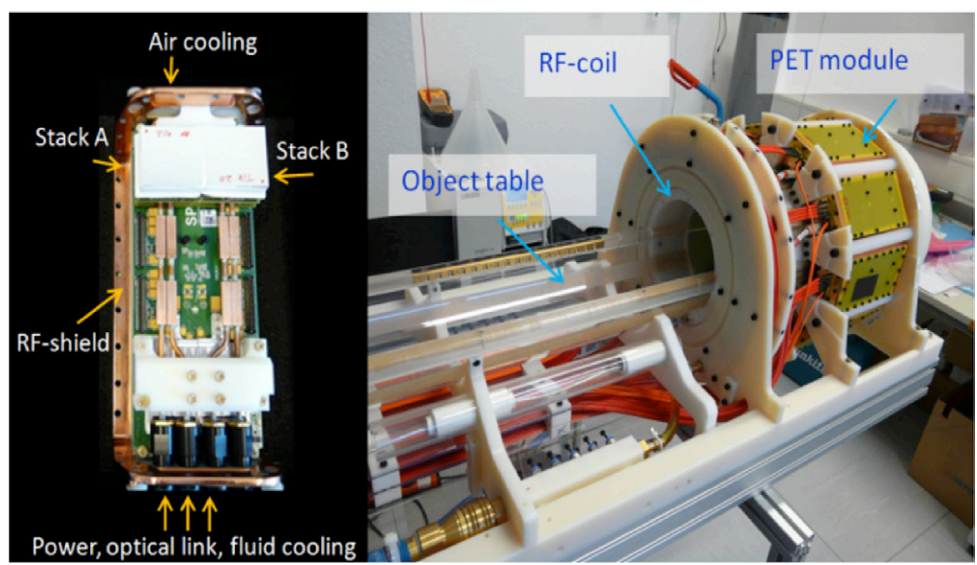

**Figure 9.** The Hyperimage module and insert for a human 3 T MRI. The system is based on analog SiPM (reproduced with permission from Schulz *et al* (2009)).

*4.2.1. APD based systems.* MRI-compatible small animal PET scanners using APDs are exemplified by those first developed at UC Davis (Catana *et al* 2006) and in Tuebingen (Judenhofer *et al* 2007, Judenhofer *et al* 2008). The Tuebingen system is shown in figure 9. The different approaches taken illustrate some of the issues encountered when detectors and electronic components are placed within the magnet bore. The PET imaging annulus is similar for both systems, consisting of a ring of LSO arrays that fit within the 11.8 icm nside diameter gradient coils a of 7 T Bruker small animal MRI system, but the readout and shielding arrangements are implemented differently.

The UC Davis scanner detects the scintillation light via 10 cm long optical fibres connected to $14 \times 14$ mm$^2$ position-sensitive APDs which are cooled by nitrogen gas to $-5$ °C and operate with a bias voltage of $-1.66$ kV. The fibres allow the APDs, preamplifiers and other readout electronics to be placed outside the RF coil and the linear regions of the gradients in the axial direction (but still within the high field of the magnet bore). This arrangement requires bending of the optical fibres so the axial PET field of view is limited to 12 mm. To exclude all conducting and magnetic components from the field of view, the APDs and electronics are enclosed in concentric cylindrical RF shields, made from a glass micro-fibre reinforced PTFE composite, however there is no shielding in the central MRI imaging volume where only scintillator, fibres and supporting structures are present. The use of position sensitive APDs in the magnet bore introduces some small distortions to the detector flood map and there is some light loss due to the bent fibres, however overall there is no significant degradation in PET or MRI performance when using standard MRI pulse sequences. In the Tuebingen (Judenhofer *et al* 2007) system the optical fibres are eliminated and each crystal array is coupled directly to a 3 × 3 APD array (each element is 5 × 5 mm$^2$) via a thin light guide, so the APDs and electronics (with non-magnetic components) are positioned within the MRI imaging volume close to the RF and gradient coils. Double-sided printed circuit board material coated with 10 $\mu$m thick copper provides an enclosure for the whole detector module—the copper layer was only 10 $\mu$m thick to avoid MRI image artefacts caused by eddy currents induced in the shielding. The system is operated at room temperature with a bias voltage of −380 V, and a detailed assessment of the various sources of interference (Wehrl *et al* 2011) showed that simultaneous PET-MRI data could be acquired for a wide range of demanding sequences including fMRI and MRIS, noting just a small reduction in MRI image SNR and a small drift in mean signal intensity (attributed to non-stable temperature control).

Maramraju *et al* (2011) describe a system with a similar overall configuration to the Tuebingen one. This system is using an APD with direct one-to-one coupling, instead of a block detector. Again, fairly small interference effects were seen, including a reduction in homogeneity of the 9.4 T magnet which was correctable by standard shimming coil adjustments. Sensitivity of the PET to RF excitation was initially addressed by switching off the PET electronics during RF excitation, resulting in deadtime of 6%–28% depending on the MRI sequence. A detailed investigation of shielding options (Maramraju *et al* 2011) demonstrated that the RF interference, and reduction of the MRI SNR due to eddy currents induced in the shielding, could be reduced to acceptable levels by using a thin segmented copper shield around the RF coil to separate it from the PET electronics, although gating of PET signals was still required for sequences with high RF power.

The systems above have demonstrated that, for most standard MRI sequences, acceptable PET and MRI performance can be obtained by placing PET detector components within the MRI field of view, provided careful attention is paid to the selection of components and shielding, and the majority of subsequent simultaneous systems have adopted this configuration. Nevertheless, the PET performance of these prototypes, particularly in terms of sensitivity which is typically <1%, lags behind that of state-of-the-art PET-only small animal systems with sensitivities of typically 4–10% . The most recent systems from these and other groups therefore feature much longer axial fields of view of 5–7 cm for whole body mouse imaging with improved sensitivity.

*4.2.2. SiPM based systems.* The European Hyperimage collaboration has constructed a fully functioning SiPM-based small animal MRI-compatible PET scanner (a picture of the system is shown in figure 10) to operate within a 3 T clinical MRI system (Schulz *et al* 2009) which allows a large transaxial field of view and provides extra capabilities for translational studies. With a ring diameter of 20 cm and axial field of view of 9 cm the system is designed to provide in a PET performance comparable with standalone systems, though the sensitivity of the first realisation of the system was limited to 0.6% as only 1/3 of the axial field of view was populated with detectors.

Each detector comprises a 22 by 22 array of 1.3 × 1.3 × 10 mm$^3$ LYSO crystals coupled by a thin light guide to an 8 by 8 array of 4 × 4 mm$^2$ SiPMs. Rather than take the approach of reducing the number of readout channels or removing electronic components from the field of view, the problems associated with a large number of readout channels from the SiPM array has been addressed head on and, in order to minimize potential crosstalk, all analogue signals coming from the SiPM elements are digitized immediately adjacent to the sensor, within the field of view, using a custom designed ASIC which outputs digital energy, timing, and channel information. The detector units, comprising the scintillator, SiPM array and ASIC, are enclosed in RF shielded module and liquid cooled to 22 °C. A custom designed transmit/

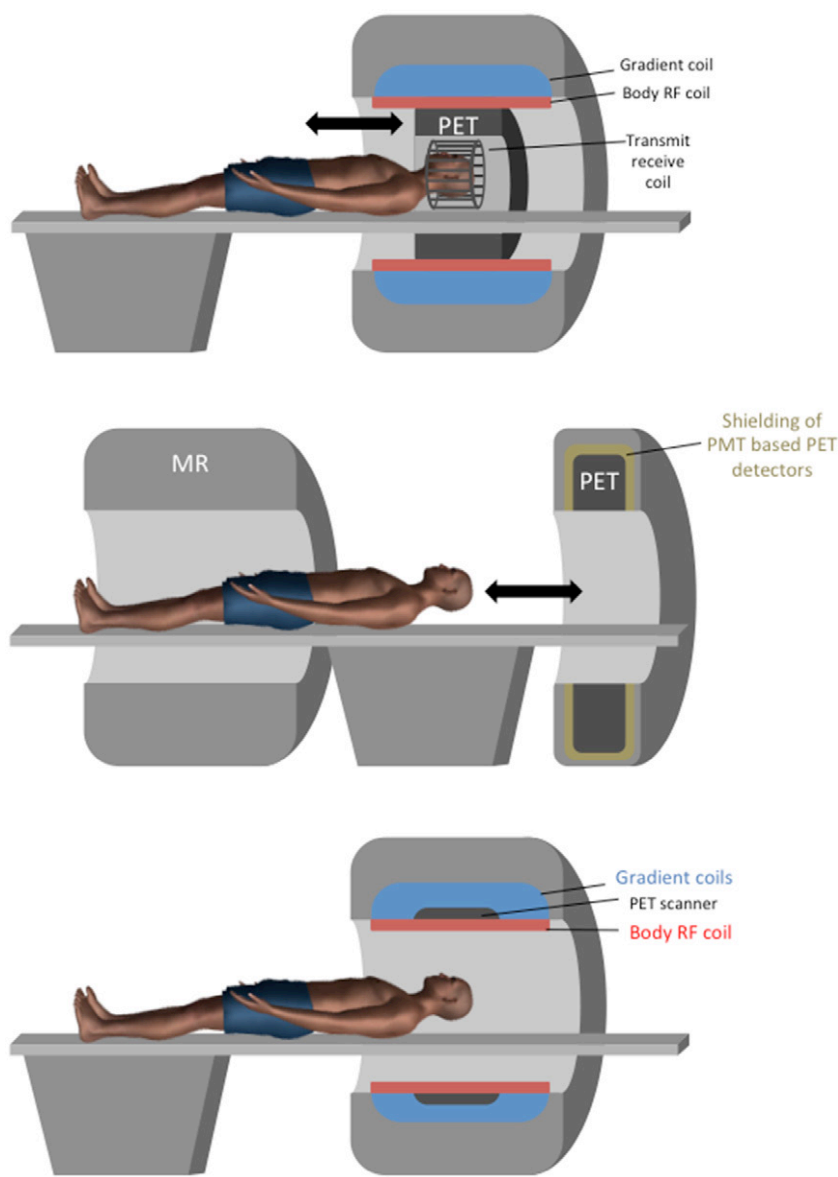

**Figure 10.** Three concepts for integration of PET and MRI: a brain PET scanner inserted in a whole body MRI (top), a sequential PET MRI scanner for whole body imaging (middle) and a fully integrated simultaneous PET MRI scanner for patient imaging (bottom).

receive coil is integrated into the gantry—the 16 cm diameter birdcage design is low density with all components moved outside the PET field of view to minimise attenuation artefacts.

Given the many potential sources of interference inherent in this approach, great attention was paid to all aspects of MRI compatibility including the selection and production of non-ferromagnetic components, differential signalling for all analogue signals, circuit layout to avoid eddy currents and heating effects and minimising RF from power supplies. These considerations are described in detail by Wehner *et al* (2014). A small increase in the MRI noise floor was measured when the PET system was operated, however overall only very minimal effects on $B_0$ and gradient homogeneity or RF related effects could be measured, and likewise no effects of RF or gradient switching on PET were measured. This is attributed primarily to the direct digitization of the SiPM-signals coupled with detailed attention to all potential compatibility issues. Although time-of-flight capability is not currently relevant for pre-clinical PET, individual detector modules demonstrated (Schulz *et al* 2009) a temporal resolution of 520 ps in a 3 T magnet indicating that this approach is also likely to be suitable for a TOF human MRI-compatible PET scanner.

A second version of the Hyperimage system has recently been constructed using digital silicon photomultiplier (dSiPM) technology (Weissler *et al* 2012). In addition to the improved spatial and temporal resolution of the detectors, the digital SiPMs are less temperature dependent, have a lower power consumption with respect to the previously used analog SiPM/ASIC combination, and as a separate ASIC to digitise the sensor outputs is not required, and the detector stack can be more compact, all of which are advantageous for MRI integration. The detector modules incorporate liquid cooling and are enclosed in carbon fibre housings which provide effective RF shielding whilst having negligible interaction with the gradients. The sensor is operated at a lower temperature than the analog in the first Hyperimage system (for details check the reference Wehner *et al* (2014)). Data transfer to and from the PET modules is all performed using specially engineered optical transceivers. The system also includes many means for synchronising data acquisition between PET and MRI, for example MRI sequence data can be inserted directly in to the PET datastream. PET performance degradation was examined under extreme MRI conditions (i.e. intense RF and fast gradient switching). Apart from some minor unresolved issues (sensitivity of the PET electronics to the z-gradients, and an increase in the MRI noise floor during PET acquisition) interactions between the two systems are negligible (Wehner *et al* 2014). MRI-PET Interference of the digital SiPM based system has recently been investigated in detail by Weissler *et al* (2014).

Several other groups have constructed small animal systems based on SiPM arrays. The basic detectors configurations are on the whole similar, however different approaches have been taken in particular to transferring signals out of the magnet and shielding arrangements. Hong *et al* (2012) place short optical fibre bundles between the crystal arrays and sensors. This arrangement results in gaps between the shielded sensor modules thus maximising RF transmission to the subject from the MRI body coil, which can then be used in conjunction with a separate receive-only coil. Kang *et al* (2011) transfer SiPM charge output signals to a preamplifier located remotely using 300 cm flexible flat cables. PET electronics can then be positioned outside of the 5 Gauss line (1.5 meters away from the magnet isocenter of the MRI. Only non-magnetic PET components, scintillation crystals and SiPM arrays without preamplifiers or subsequent electrical circuits need be placed inside the MRI bore, and no electromagnetic shielding was used to protect the PET components. This group has used the same concept to construct a human brain system(see below)

### 4.3. Brain insert for simultaneous imaging

The first simultaneous human brain imaging PET-MRI systems was described in 2007, around the same time as the first descriptions of APD systems for small animal imaging. This PET scanner is inserted in the bore of the MRI and inside the PET scanner a RF coil for brain imaging is positioned (as shown in figure 3).

A limited number of the Siemens brainPET insert (Kolb *et al* 2012) were constructed as an experimental clinical system with a high resolution PET insert that could be inserted into (and removed from) the bore of a Siemens 3 T MRI system. The overall system configuration is shown schematically in figure 3 (top), and the specific and the specific system design and setup are shown in figure 11. LSO/APD based PET detectors, similar in principle to those used in the Tuebingen small animal system described in the section on APD small animal systems above, were used and the device allowed many practical and technical issues to be evaluated and resolved prior to the subsequent incorporation of similar technology into a whole body simultaneous PET-MRI system (Delso *et al* 2011). The PET detector module itself comprises a fine LSO array (12 × 12 array of 2.5 × 2.5 × 20 mm$^3$ crystals) coupled via a short light guide to a 3 × 3 array of 5 × 5 mm$^2$ APDs. The detectors and adjacent front end electronics are air-cooled and contained within 32 copper-shielded detector cassettes and the insert is placed inside the MRI body coil . As this is a brain imaging system the body coil is disabled when the PET is inserted, and a quadrature transmit/receive circular-polarised head coil, designed to minimise gamma photon attenuation, is placed inside the PET, fixed to the patient couch, within the PET field of view.

Because APD photodetectors are employed, the coincidence resolving time is a relatively long 4.9 ns—this is not fast enough for time of flight imaging, but no current PET systems have a timing resolution fast enough to be relevant for brain imaging. The small diameter (37.6 cm) and a relatively long axial extent (19.1 cm) result in a very high sensitivity of 7%, which combined with a reconstructed spatial resolution of <3 mm FWHM results in excellent image quality, however, it took considerable time and effort to address the many artefacts initially present in reconstructed images including effects due to the structure of the PET ring (gaps are present between the PET modules), the effect of out of field of view activity (no end shields) and the presence of the RF coil in the field of view. Many of these effects were addressed for the first time in this system giving a foretaste of the many difficulties to be encountered in obtaining accurate PET quantification in PET-MRI.

It was shown that for standard sequences interference between the two systems in terms of $B_0$ and $B_1$ homogeneity was minimal. The ability to perform complex functional studies such as functional MRI (fMRI) and proton spectroscopy was demonstrated. A small increase in the MRI noise floor to PET RF interference was found, and in order to determine the cause of the count rate reduction, Weirich *et al* (2012) performed a detailed investigation that illustrates the complexity of the system. An instantaneous reduction of the PET sensitivity was observed for MRI sequences with fast switching gradients such as echo planar imaging (EPI). The RF pulses, the switched gradient fields and constant magnetic field and detector temperature were all examined. A significant count rate reduction (Weirich *et al* 2012) up to 3% was observed in standard sequences and a correction method depending on the gradient amplitude has been designed, implemented and evaluated. The conclusion was that the effects related to PET reconstruction and calibration are more significant than effects due to direct PET-MRI interference.

A brainPET PET insert, nearly identical to that used with the 3 T scanner above, has also been installed in an 'ultra high field' 9.4 T human brain MRI system, with a bore diameter of 90 cm, at Juelich (Shah *et al* 2013). In addition to structural imaging at significantly increased spatial resolution, such a high field strength provides the possibility of non-proton MRI and spectroscopy with a spatial resolution comparable to PET, however various complications with high-resolution, proton-based imaging MRI at this high field still need to be addressed before complementary PET-MRI studies can begin.

The only other brain system reported to date is the prototype system from Hong *et al* (2013). Again this operates in a 3 T magnet. It follows the same general principles as the pre-clinical system from the same group described above. An annulus of LYSO/SiPM arrays is situated within the bore of the MRI, but all other electronics is located outside the magnet just as for the preclinical system above, signals being transmitted directly from the SiPMs using long flat shielded cables. Simultaneous imaging of phantoms has been demonstrated, although small degradations of some PET parameters are seen due to RF noise pickup. Larger SNR decreases on the MRI are attributed to eddy currents in the unsegmented shielding material of the PET gantry.

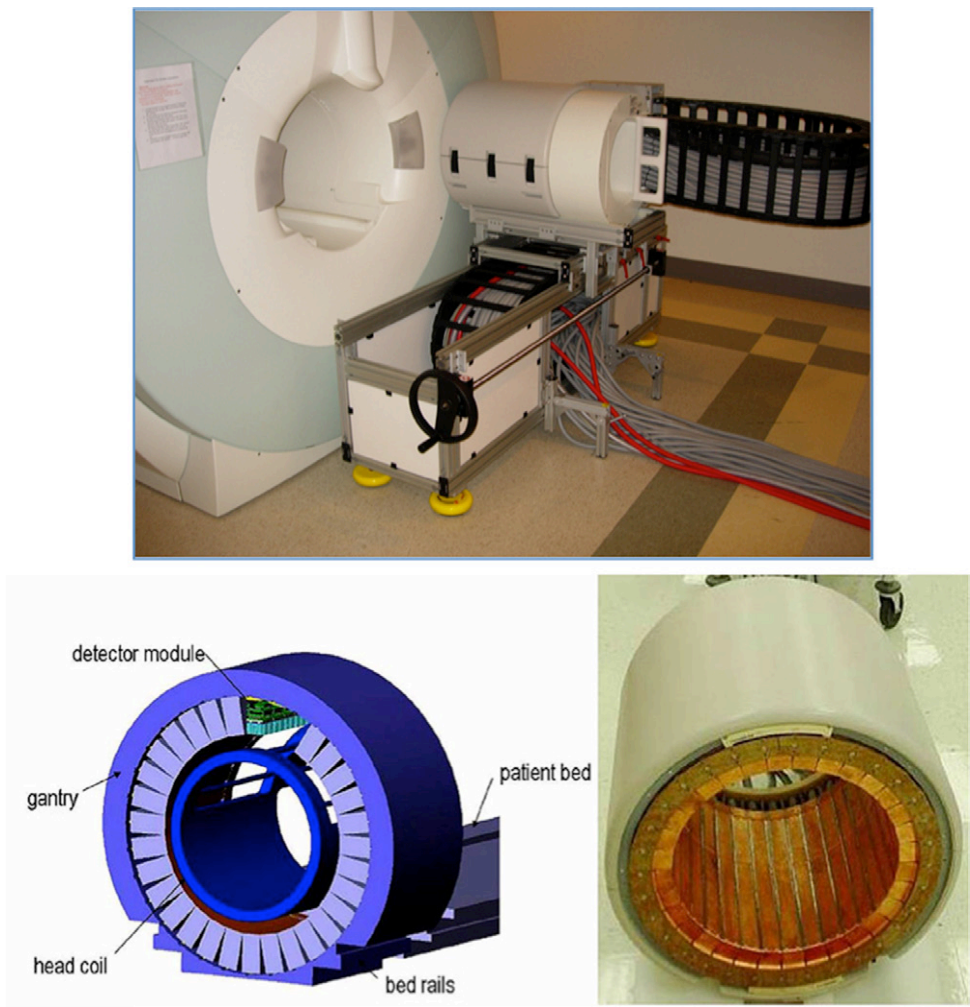

**Figure 11.** The brain PET insert is based on APD detectors and can be positioned inside a Siemens 3 T MRI scanner. Reproduced with permission from Cherry *et al* (2008).

### 4.4. Sequential whole body systems

The sequential scanner configuration (as shown in figure 3) cannot be used for investigations that require temporal correlation of PET and MRI and also results in a longer total acquisition time than simultaneous acquisition. However the technical challenges are much simpler, so this is an attractive approach for exploring potential clinical PET-MRI applications, and in particular MRIAC, without having to establish a fully functioning clinical simultaneous system. This is one of the stated aims of the tri-modality sequential system from GE Healthcare (Veit-Haibach *et al* 2013) which, as its name suggests, incorporates a standard fully featured MRI and PET-CT in separate, ideally adjacent, rooms and a purpose designed trolley system to transfer the patient between the two. As the patient remains on the same couch-top throughout, patient movement is kept to a minimum. An advantage of this arrangement is that the investigation of MRIAC can be performed with or without MRI coils in the PET field of view—if the MRI is performed first the coils must be carefully removed prior to the PET-CT.

The Ingenuity TF sequential PET-MRI from Philips Healthcare (Griesmer *et al* 2010, Zaidi *et al* 2011) was the first commercially available human PET-MRI system. It comprises an unmodified 3 T whole body MRI system and a modified whole body ToF PET scanner facing each other with just 4.2 m separating the MRI and PET fields-of-view. A sophisticated sliding couch arrangement (no trolley is involved) allows the patient to have a whole body scan first in the MRI and then in the PET with minimal transfer time. As the two systems are relatively close, the majority of the PET electronics is situated outside of the scan room to minimise RF interference, and local magnetic shielding reduces the fringe field strength to ∼0.1 mT around the PET detectors. The PMTs are also realigned to minimise loss of gain and the PMT high voltage is reduced from between 1100 and 1500 V dc to 900 V dc while acquiring MRI data. With these measures in place there is essentially zero interference between the two systems. The only other modification to the PET scanner is that there is no CT scanner or other transmission source so it is totally dependent on the MRI scanner for attenuation correction which is achieved via a segmented optimised anatomical MRI scan (see discussion of MRIAC techniques in section below). As for the GE system above, it is possible to remove anterior coils prior to the PET scan, but pre-generated attenuation maps of rigid coils, located in pre-determined positions, and other accessories are available if these items are in the field-of-view during the PET acquisition.

The system decribed by Cho *et al* (2007) has a different rationale to those above. The aim is to obtain very high resolution, very accurately spatially but not temporally co-registered PET and MRI images of the brain by combining very high specification PET and MRI systems. An unmodified Siemens HRRT PET (PMT based brain PET with 2.4 mm FWHM spatial resolution) and a 7 T MRI (200 $\mu$m FWHM) are installed facing one another and 6 m apart such that the subject can be moved between the two using a highly accurate shuttle system. In order to achieve a magnetic field of less than 0.05 mT at the PET location, the MRI was located in a magnetically shielded room and it was necessary to place steel shielding around the PET scanner (total 500 tons of steel).

An issue for all these systems (and to a lesser extent for simultaneous systems, though to what extent remains to be seen) is exactly how accurate the PET-MRI registration is, how accurate does it need to be, and can it be improved if necessary by image registration techniques. For the GE tri-modality system, Samarin *et al* (2011) measured the mean offsets between CT and MRI in the abdomen (following a rigid manual image registration procedure using anatomical landmarks) to be 1.5, 0.9 and 4.4 mm in x, y and z respectively. This is comparable with the 3.9 mm (RMS) obtained for patient registration accuracy between the PET and CT components of a PET-CT scan and better than the 8.5 mm registration accuracy for PET-CT to MRI registration reported by Somer *et al* (2007).

*4.5. Simultaneous whole body systems*

Based on the experience gained with the brainPET system the Siemens mMR whole body integrated simultaneous PET-MRI (Delso *et al* 2011) was developed and is the first commercially available clinical whole body simultaneous PET-MRI (shown in figure 12). The PET detector assembly is installed between the gradient and body coils of a 3 T whole body MRI. The system is composed of 8 rings of 56 detector blocks. Per block there are 8 × 8 LSO crystals (4 × 4 × 20 mm), coupled to an array of 3 × 3 APDs (water-cooled), totalling 4032 channels. The temporal resolution is 2.93 ns, so again the system does not have time-of-flight capability. Due to the relatively small PET ring diameter (transaxial field of view: 59.4 cm) required to fit within the MRI bore, large axial FOV (25.8 cm) and large coincidence window an increase in random and scattered counts might be expected, however, this seems to be compensated for by the narrower energy-window settings, which, in combination with comparable energy resolution, lead to a noise-equivalent count rate better than most PET-CT systems and a good scatter fraction, indicating that the integration of the PET detectors in the MRI scanner and their operation within the magnetic field does not have a perceptible impact on the overall performance. The MRI subsystem performs essentially like a standalone system, however, further work is necessary to evaluate the more advanced MRI applications, such as functional imaging and spectroscopy.

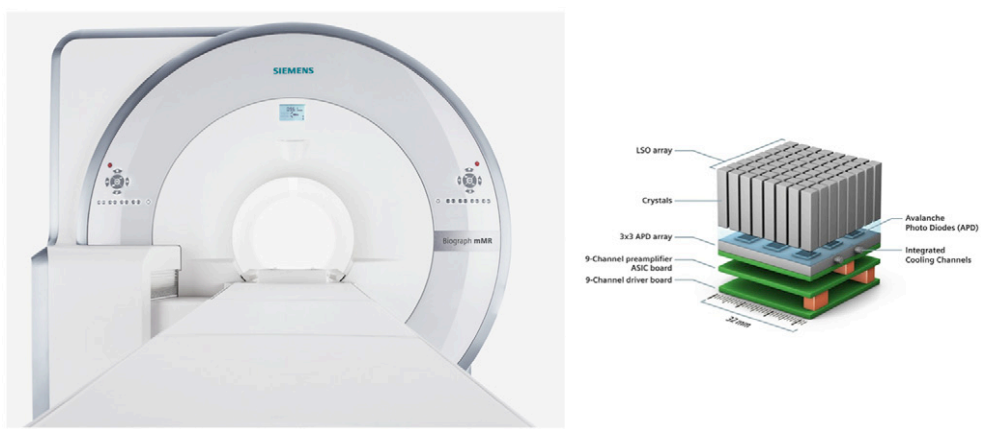

**Figure 12.** The Siemens mMR system (Delso *et al* 2011) is based on a compact APD PET detector (source of material: www.healthcare.siemens.com).

GE recently announced their intent to market a clinical whole body simultaneous TOF-PET-MRI system (shown in figure 13). The three first versions of the system are installed in Zurich, Stanford and UCSF. The system has been presented in detail at the latest PSMRI conference in Kos. The following specifications were given during this presentations, these are described in the section below.

The MRI component is based on the GE 3 T Discovery 750 w MRI system which has an inner bore of 70 cm diameter. The first description of this Time-of-flight PET system for PET-MRI has been published in Levin *et al* (2013). The TOF-PET has a transverse FOV of 60 cm and an axial extent of 25 cm (89 slices). The total detector thickness (including electronics and cooling) is less than 5 cm. After integration the PET-MRI system has a bore of 60 cm and the full performance of the MRI system is maintained after the integration with the PET. The detector is based on an Lutetium Based scintillator (LBS), with similar density as L(Y)SO. The scintillator dimensions are 4 mm in the transverse direction, 5.3 mm in the axial direction and the thickness is 25 mm. Readout is done with a pixelated analog SiPM in combination with limited light sharing (UV transparent light guide). Light sharing enables to keep the number of readout channels manageable. There is a 2 fold reduction in electronic channel counts which allows a design with low power requirements. Readout is done by a custom made ASIC, this permits to adjust the gain at the level of the individual SiPM pixel. In combination with temperature sensors gain stability is obtained during a range of thermal loads (Kim *et al* 2014).

The reported specifications (Levin *et al* 2014) of the system are 10.5% energy resolution and 390 ps average timing resolution for the whole system. This is superior time of flight performance than current clinical PET-CT systems and in the same range as the Philips Vereos PET-CT based on digital SiPMs.

The transverse spatial resolution is 4.2 mm. The PET NEMA sensitivity is 22.5 kcps MBq$^{-1}$ with a line source at the center. The high sensitivity is obtained by the combination of large detector thickness, small detector bore (62 cm face-to-face) and long axial FOV. Additionally sensitivity is enhanced by recovering annihilation photons that interacted by Compton scatter inside the PET detector (Wagadarikar and Dolinsky 2014). The peak NECR is 215 kcps at 17 kBq ml$^{-1}$ activity. Comparable results were obtained for the PET NEMA IQ phantom with and without the RF of the MRI turned on.

Attenuation correction is under development and evaluation and will be based on MRI and/or TOF-MLAA based algorithms (see section on attenuation correction for more detail on these algorithms). The main motivation is that several studies have shown the smaller effect of artefacts in attenuation map when TOF information is used during emission reconstruction.

## 5. Quantitative image reconstruction in PET-MRI

Besides the integration of the hardware components, there is also a need for reconstruction algorithms with the necessary corrections. As mentioned before one of the important advantages of PET-CT compared to standalone PET is the fast acquisition and easy transformation of CT data into attenuation correction factors for PET image reconstruction. In PET-MRI the attenuation correction seems to be one of the major barriers to the full acceptance of PET-MRI as an established clinical imaging modality. The different methods investigated to overcome this limitation are described in this section.

### *5.1. Attenuation correction for PET-MRI*

*5.1.1. Requirements for attenuation correction.* The major image degrading factors in PET are the attenuation and scatter from the object inside the FOV (Kinahan *et al* 1998, Turkington 2000). The reconstruction of quantitative PET images requires accurate correction for attenuation of 511 kev gamma rays, which is a large effect. The half value layer thickness for a 511 keV gamma ray in tissue is about 7 cm. To detect a coincidence both photons should not be attenuated and therefore the fraction of gamma ray pairs attenuated from the inner parts of the body can be higher than 90 percent. An accurate correction method is therefore essential both to avoid image distortions and artefacts and to permit accurate regional quantification both for routine clinical studies and for quantitative dynamic studies. If a map of linear attenuation coefficients at 511 keV in the object is available then it is straightforward to perform an attenuation correction. Such a map can be obtained rapidly (in a couple of minutes for a whole body) using the CT scanner component in PET-CT (Kinahan *et al* 2003), however none of the integrated PET-MRI systems constructed to date incorporate a CT scanner or rotating transmission source, so this information must be somehow derived (Martinez-Möller and Nekolla 2012) from the MRI image and/or any other available sources of information on attenuation like additional sources or PET emission data.

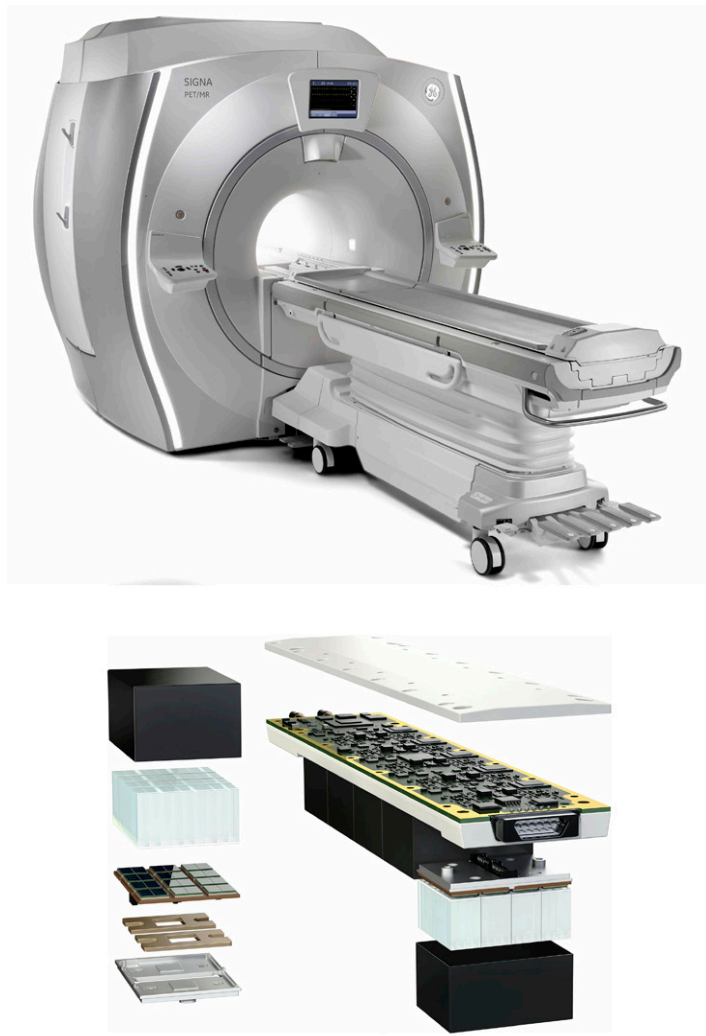

**Figure 13.** The GE PET-MRI system (Levin *et al* 2013) (picture on top) is based on a compact LBS based analog SiPM PET detector (picture at bottom) (reproduced with permission from GE Healthcare).

The effect of attenuation and the requirements for the accuracy of the attenuation map depends strongly on the object size. For small animal imaging the effect of attenuation is much smaller and methods like contour based attenuation correction derived from MRI will have sufficient accuracy for most studies (Keereman *et al* 2012). In this part we therefore only focus on the problem of attenuation correction for human brain and whole body imaging. Three different groups of attenuation correction techniques are currently being investigated for PET-MRI: MRI based, emission based and transmission based (Bezrukov *et al* 2013, Keereman *et al* 2013). Within each of these groups several methods with differences in acquisition and data postprocessing are under investigation as illustrated in figure 14 for the majority of studies CT based attenuation correction is used as a reference for comparison.

Before going into more detail for these different approaches, it is important to determine the requirements of the attenuation correction method.

The main requirement for the implementation of an attenuation correction method in clinical practice is the robustness of the method, i.e. the guarantee that no error that could lead to an incorrect diagnosis will be present in the attenuation map. To enable this, it is required that all attenuating objects (also the ones on the path between the emitting object and detectors) inside the FOV are included in the attenuation measurement. A specific problem for PET-MRI is the attenuation caused by MRI coils and other hardware within the field of view typically positioned close to the object of interest. Such objects contribute to the total attenuation, thus reducing the PET signal, and can result in artefacts if not accurately corrected for.

Any attenuation correction method should also have a reasonable acquisition time: ideally the attenuation map is acquired simultaneously with the PET data (to avoid misalignment) and it should only lead to a minimal increase of the total acquisition time and very limited additional dose to the patient.

Several of the attenuation correction techniques involve a segmentation step. A detailed simulation study by Keereman *et al* (2011) has determined the number of tissue classes that needs to be segmented to have acceptable errors in the emission reconstruction. Air, lung, soft tissue, spongeous bone and cortical bone should be present, while adipose tissue is not required. There is a large variation in the attenuation coefficients in the lung for different patients (Keereman *et al* 2008). As this is a large volume in the body, the patient specific linear attenuation coefficients should be obtained to avoid large quantification errors. (Martinez-Möeller *et al* 2009, Steinberg *et al* 2010 and Eiber *et al* 2011)

Specific requirements will also depend on the area of interest. For head scans detection of lung tissue is not needed but the relative amount of skull to soft tissue will be relatively high. Truncation of arms is not relevant in this case. For torso scans accurate detection of the lungs is required and truncation of the FOV can have a significant effect.

*5.1.2. MRI-based attenuation correction.* There are several major challenges in the development of an attenuation correction method based on MRI data.

(a) MRI image intensity values reflect proton density and tissue relaxation properties (Stanisz *et al* 2005) so there is no direct relationship between the signal measured in MRI and the linear attenuation coefficient at 511 keV. MRI based methods are therefore based on segmentation of the object into different tissue classes which are assigned a predefined linear attenuation coefficient.
(b) Standard MRI sequences give a very low signal in bones (which is the most attenuating tissue in the body) and lung because of the low proton density and the very short $T_2$ of these tissues (Stanisz *et al* 2005). Also air has a very low signal in these sequences. Therefore these different tissues can not be differentiated based on the MRI signal (from standard sequences), whilst their respective gamma attenuation coefficients are very different.
(c) The field-of-view of the MRI system is much smaller than the field-of-view in PET. This can lead to truncation of arms in patients.
(d) There is a wide variety in the size, shape and composition of the attenuating objects that can be present in the FOV of the PET-MRI system and these can cause significant attenuation (Delso *et al* 2010). Therefore it is difficult to make prior assumptions about the object. The advantages and disadvantages of each method are summarized in table 3.

The only potential advantage of MRI based attenuation correction compared to PET-CT is the simultaneous acquisition. In a PET-CT scanner there is a mismatch in acquisition time between the PET emission data (typically acquired as an average image over many respiratory cycles) and CT (which provides a snapshot at one phase of the respiratory cycle). This introduces artefacts at the boundary between the lungs and diaphragm (Kinahan *et al* 2003). Artefacts may also result from the use of CT contrast agents and in the presence of metallic implants and prosthesis, and PET-CT registration may not be perfect. In a simultaneous system the acquisition of PET and MRI data can be acquired during the same breathing phase, which can minimise these artefacts.

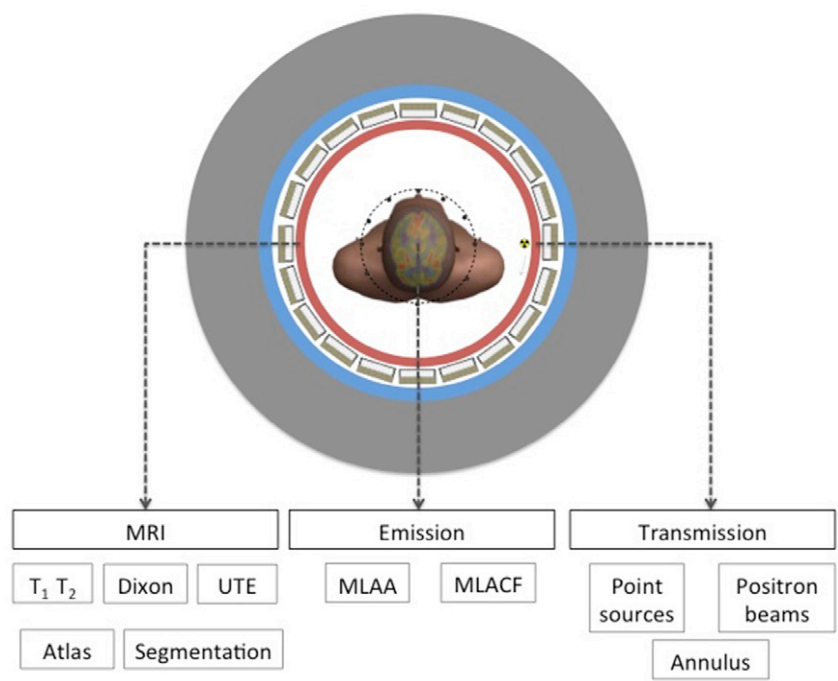

**Figure 14.** Different sources of information to derive attenuation correction maps in PET-MRI.

**Table 3.** Performance of methods for attenuation correction in PET-MRI.

| Method | Advantages | Disadvantages |
| --- | --- | --- |
| MRI(Segmentation) | Fast<br>No dose | Segmentation errors<br>No signal in bone<br>One AC value per tissue<br>Need for templates for coils<br>Truncated FOV |
| MRI(Atlas) | Fast<br>No dose | Anatomical abnormalities<br>Difficult for body imaging<br>Templates for coils<br>Truncated FOV |
| MRI(UTE) | Identification of bone | Additional MRI acquisition time needed<br>Not tested for whole body imaging |
| PET(Emission) | No additional acquisition time | Limited to tracers with distributed uptake (like FDG)<br>Need for templates for coils |
| PET(Transmission) | Works for any object in FOV | Additional sources and dose<br>Noisy attenuation maps<br>Limited spatial resolution |

*Segmentation based MRI-AC.* In segmentation approaches, one or more MRI images are acquired and segmented (Berker *et al* 2012) into discrete anatomical regions. The ability to identify different regions and the number of tissue classes to which these can be assigned largely reflects the MRI sequences used. The regions are labelled on the basis of their MRI signal intensity or their anatomical location (e.g. the shape of the lungs can be identified), and pre-defined attenuation values are assigned to them. In addition to the limitations of the available MRI images, the main sources of error and inaccuracy are the segmentation procedure, misclassification of regions (e.g. because bone is not accounted for) and the inability to account for attenuation values that vary from subject to subject (e.g. in the lungs) (Keereman *et al* 2013). Segmentation can be performed with many techniques ranging from simple thresholding and morphological operations to more complex approaches that make use of a priori knowledge to identify, for example, different structures in the brain. he main advantage of the segmentation approach is that, in principle, it can account for the great variability in the shapes and positions of organs in the body. The main disadvantage is that in practice not all tissues and structures can be reliably identified and it is difficult to account for subject-specific and local variations in $\mu$-values. There are some efforts to find a relationship between the $\mu$-values and the MRI signals. This can account for inter-patient variability to some extent but this has proved very difficult. T

In order to distinguish adipose- and water-based tissue the Dixon sequence can be used (Dixon 1984). As the Larmor frequency of protons in water and fat is different, one can distinguish water and fat from MRI images acquired at different echo times (Eiber *et al* 2011). It has been demonstrated that this sequence is also able to serve as an anatomical scan for localisation of PET uptake. Efforts have been made to extract the lungs from these images using segmentation of areas with very low signal.

The Dixon based and other standard MRI methods do not allow detection of bones. Ultrashort EchoTime (UTE) sequences are able to visualise tissues with very short $T_2$ relaxation time (Catana *et al* 2010, Keereman *et al* 2010). These methods can provide images where bone is shown with sufficiently high signal. The combination of images acquired at different echo times allows it to be differentiated from air. The initial acquisition times used in UTE sequences are quite long, typically 3–5 min per bed position and for larger objects (body region) artefacts appear in the acquisition. Therefore up till now UTE sequences only seem suitable for brain imaging and not for multi-position whole body imaging.

*Template and atlas based MRI-AC.* One way to circumvent the need for more complex or lengthy acquisition sequences is to make use of knowledge-based and machine-learning methods (Hofmann *et al* 2008). Atlas based methods are based on a number of registered MRI-CT datasets, forming the atlas. All MRI images of the atlas are then registered to the measured patient MRI image. In atlas-based approaches a $\mu$-map or pseudo-CT template (Kops and Herzog 2013) in a standard anatomical space is created and then transformed to fit an anatomical MRI of the patient. The template can be derived in any number of ways, the most obvious being some combination of CT scans or PET transmission scans acquired from a large number of subjects. The template can be a simple average of a large number of CT scans or it could be a more complex database of scans where the most appropriate components are selected on the basis of the detailed structure of the individual's MRI. The potential advantage of this technique is that the template can incorporate much more structural detail than can be segmented and labelled in the approaches described above. In particular bone could be included without requiring UTE acquisitions. It can also contain a continuous range of $\mu$-values which will be attached to the correct anatomical regions. Whilst the method may be effective for the brain, accounting for all the anatomical variation in the body will be very difficult. The applicability will also remain limited to a population without anatomical abnormalities.

*Evaluation of MRI-AC on patient data.*   The first evaluations of the accuracy of the different methods have been performed using data acquired on standalone PET-CT and MRI systems or the Siemens brain PET-insert. For brain imaging good results have been obtained using template and atlas based methods. UTE methods (Catana *et al* 2010, Keereeman *et al* 2010) have been shown to further improve the accuracy due to the more accurate accounting for the density of bones. More detailed results can be found in review papers on the topic of attenuation correction.

More recently evaluation of the different techniques on clinical data acquired on simultaneous PET-MRI has begun (Hofmann *et al* 2011). As the attenuation correction method implemented on most clinical PET-MRI systems is based on Dixon sequences, most clinical evaluations are concerned with this technique. As no gold standard is available data are typically compared to PET-CT of the same patient, acquired earlier, However such a comparison can be confounded by the potential changes in SUV due to different acquisition times after injection.

*5.1.3. Attenuation correction based on data from PET.*   An alternative for MRI based attenuation correction is to derive the attenuation information from the data collected by the PET scanner. As the acquisition time in PET is shorter than the typical acquisition time of the MRI part, this method is interesting for patient throughput. Several methods have been developed that rely on emission and/or transmission data. The main difficulties of transmission and emission based methods are clearly different from MRI based methods.

*Transmission based attenuation correction.*   Transmission imaging has been used extensively in standalone PET scanners and has a clear advantage because it results directly in the linear attenuation coefficient at 511 keV (or a nearby energy for singles transmission). Another advantage (compared to MRI based methods) is its general applicability: with this method it is possible to determine the linear attenuation coefficient of any object (also coils) inserted in the FOV. There are however also some challenges in the development of an attenuation correction method based on transmission data for PET-MRI.

 (a) Transmission images can only be acquired with limited statistics due to dose constraints and the limited countrate of the PET system. This results in noisy transmission images with limited spatial resolution
 (b) In an integrated whole body PET-MRI the bore is small compared to a standard PET scanner and limited space is left for introduction of a transmission source. Due to the magnetic field rotating sources may also be difficult to implement
 (c) Acquisition times are long for sequential emission/transmission imaging

(d) In simultaneous emission/transmission scans crosstalk will be present between transmission and emission data

To obtain an acceptable time for the transmission measurement one has to develop either a very fast sequential transmission scan (difficult with PET detectors due to count rate capability) or collect emission and transmission data simultaneously. Using TOF information it becomes possible to separate emission from transmission data if the distance from emission to the transmission source is sufficiently large. This approach has been validated using simulations and an experimental setup based on an annulus transmission source (Mollet *et al* 2012). It was shown that sufficient statistics could be obtained to derive attenuation maps using an iterative reconstruction method. The main challenge in this approach remains the crosstalk and optimisation of corrections for scatter and count rate effects. By using another source geometry (fixed number of line sources) these contamination effects can be minimized. The method is however only applicable to systems with TOF information, like the sequential Philips TOF PET-MRI (Mollet *et al* 2014) or the recently introduced GE TOF-PET-MRI. Most current clinical human PET-MRI scanners are however based on APD detectors which are not capable of measuring TOF differences. The use of scatter information to derive the attenuation map has recently been proposed by the group in Aachen (Berker *et al* 2014).

A novel transmission source measurement has recently been proposed by Watson *et al* (2013): this technique is making use of the main magnetic field in an MRI to inject positron beams from unshielded external emission sources. These beams can be stopped by material inside the FOV and used as transmission sources. This method is interesting for PET-MRI as it makes it possible to design a compact transmission source.

*Emission based attenuation correction.*　　These methods are based on algorithms like Maximum Likelihood reconstruction of Attenuation and Activity (MLAA) developed prior to the development of integrated PET-MRI systems. By exploiting the consistency conditions which must be satisfied by the non-attenuated data one can derive the attenuation sinogram directly from the emission data (Nuyts *et al* 1999). It has been shown that the emission data are not sufficient for deriving the attenuation correction factor. It was recently shown that when the TOF information is available a unique solution (up to a constant (Defrise *et al* 2012)) can be found for attenuation and emission (Rezaei *et al* 2012). The methods have been adapted to PET-MRI acquisitions.

In the case of truncated data a large part of the attenuation map is already known (assuming a correctly determined MRI based attenuation map). The available emission data is used as input for solving the missing data. To determine the truncated part of the attenuation map a modified MLAA algorithm can be used. The method has shown to be able to determine a good estimate of the truncated arms and shoulders.

In general the emission based techniques also have important limitations:

(a) To derive the attenuation coefficients of different regions inside the body, activity should be present in the different regions. While this is true for tracers like FDG, other tracers may not have uptake in all areas.
(b) Attenuating objects outside the patient do not have emission
(c) The complexity of scatter correction increases and may result in crosstalk between the estimation of emission and attenuation

*5.1.4. Future developments in AC for PET-MRI.*　　A wide variety of methods have been developed and tested for solving the attenuation correction problem in PET-MRI. Significant progress has been made but none of the standalone methods is currently able to achieve the same accuracy as CT based attenuation correction in PET-CT. MRI based methods seem to fail for body scans in a relative large number of patient studies due to artefacts and errors in segmentation while transmission based methods deliver attenuation maps of lower quality than CT. Potentially improved methods can be developed in the future by combining sources of information. Salomon *et al* (2011) used the emission based technique to determine the attenuation coefficients from different anatomical regions. These regions can be derived from MRI images but also have the limitations of data truncation and segmentation errors. A recent paper (Panin *et al* 2013) already combined the emission data with a simultaneously acquired external transmission source. The updates of the emission and attenuation image are alternated and it was shown that use of an external source improves the reconstruction of attenuation coefficients from emission data.

*5.2. Improvements in image reconstruction for PET-MRI*

PET-MRI also offers potential benefits for the improvement of PET images. High resolution anatomical information and motion information from MRI can be used to enhance PET image quality.

*5.2.1. Resolution modelling and improvement.* Several methods have been proposed to correct for the partial volume effect in PET (Rousset *et al* 2007). PET reconstruction can be modified and improved with prior information derived from MRI (Müller-Gärtner 1992). The collection of co-registered MRI and PET data is easier in a simultaneous system and especially for brain imaging resolution enhancement can result in nice high resolution PET datasets (Vunckx *et al* 2012). A recent overview of the different methods to improve PET image reconstruction using MRI information was given in Bai *et al* (2013). Compared to PET-CT image reconstruction however one also needs to account for the effect of the magnetic field on positron annihilation distribution (Kraus *et al* 2012). The positron distribution will be highly non-isotropic due to the main magnetic field. This can introduce effects at the boundaries of tissues with low and high density (ag. at edges of lungs).

*5.2.2. Motion correction.* Due to the extended time required to acquire PET images, images of many parts of the body are subject to the effects of patient motion. This may be due to periodic motion, as in the case of the lungs or heart, or non periodic motion as in the case of peristalis of the GI tract, or bulk patient movements during the duration of the PET acquisition. The effects of blurring due to patient motion during the PET acquisition fall into two broad categories—firstly, for small regions of activity blurring of the PET image leads to a reduction in apparent activity due to an effect similar to the partial volume effect, thus reducing quantitative accuracy. Secondly, the reduction in apparent activity for small regions of uptake may result in sufficient reduction in contrast that such regions can not be detected in a noisy background, resulting in reduced sensitivity for detection of small lesions.

　　External devices (belts, triggering, optical cameras) have been used to derive input for motion correction (Rahmim *et al* 2007), but motion of the internal organs can only be derived with limited accuracy. Attempts to correct motion using PET data alone is only successfull in regions with high enough uptake to yield sufficient motion information. Motion detection for tiny structures, e.g. arteriosclerotic plaques or small tumors, can prove difficult.

　　PET-MRI offers the possibility of acquiring high quality anatomical images continuously throughout the PET acquisition at a high dynamic rate, exactly spatially and temporally correlated to the PET emission data. In principle a similar task can be attempted with CT however there are severe limitations of radiation dose, and the acquisition is not simultaneous. The development of MRI based motion correction (Ouyang *et al* 2013) involves several steps. Firstly, dedicated MRI acquisition protocols are required for sufficiently rapid dynamic imaging sampling of the field of view. In a next step motion fields need to be extracted from the measured data and finally one needs to apply these to the PET data in such a way that a single motion-free PET image can be obtained. The motion fields can be applied either to a series of reconstructed PET images ('reconstruct-transform-average'), or can be applied as part of the reconstruction process itself (Polycarpou *et al* 2012). Whilst evidence of the effectiveness of MRI-based motion PET correction is only just starting to emerge, it has been suggested that this may be essential in order realise the higher resolution promised in the next generation of PET systems (Polycarpou *et al* 2014).

## 6. Conclusions

Since the initial idea of combining PET and MRI significant efforts have been put into the development of sequential and simultaneous PET-MRI systems. The development of MRI compatible detectors based on compact solid state photomultipliers has recently led to complete simultaneous fully integrated PET-MRI systems for human imaging. These systems are now installed in several clinical centres worldwide and a large number of researchers are investigating the use of this promising modality. One of the remaining limiting factors in the use of these systems is the attenuation correction which does not yet have the same accuracy as the one on clinical PET-CT. MRI imaging is now also used to guide image reconstruction and to perform motion compensation in order to improve the image quality of PET. The focus of PET-MRI research is now reoriented to the operational and quantitative aspects of (in particular, human) PET-MRI systems and to the development of synergistic combinations of multifunctional MRI sequences and PET tracer protocols which we believe will be the key to realising the added value of simultaneous PET and MRI.

## Acknowledgments

The authors would like to thank P Mollet for providing input for the pictures on the PET-MRI systems designs and V Keereman for providing input on the section on attenuation correction and K McNamara for administrative support. The authors from this paper have received funding from the EU FP7 projects HYPERimage (grant agreement 201651) and SUBLIMA (grant agreement 241711). We also acknowledge financial support from the Department of Health via the National Institute for Health Research (NIHR) Comprehensive Biomedical Research Centre award to Guy's and St Thomas' NHS Foundation Trust in partnership with King's College London and King's College Hospital NHS Foundation Trust. The views expressed are those of the authors and not necessarily those of the NHS, the NIHR or the Department of Health.

## References

Bai B, Li Q and Leahy R M 2013 Magnetic resonance-guided positron emission tomography image reconstruction *Semin. Nucl. Med.* **43** 30–44
Berker Y *et al* 2012 MRI-based attenuation correction for hybrid PET/MRI systems: a 4-class tissue segmentation technique using a combined ultrashort-echo-time/dixon MRI sequence *J. Nucl. Med.* **53** 796–804

Berker Y, Kiessling F and Schulz V 2014 Scattered PET data for attenuation-map reconstruction in PET/MRI *Med. Phys.* **41** 102502

Beyer T and Townsend D W 2006 Putting 'clear' into nuclear medicine: a decade of PET/CT development *Eur. J. Nucl. Med. Mol. Imag.* **33** 857–61

Bezrukov I, Mantlik F, Schmidt H, Schölkopf B and Pichler B J 2013 MR-based PET attenuation correction for PET/MR imaging *Semin. Nucl. Med.* **43** 45–59

Bindseil G A, Gilbert K M, Scholl T J, Handler W B and Chronik B A 2011 First image from a combined positron emission tomography and field-cycled MRI system *Magn. Reson. Med.* **66** 301–5

Blamire A M 2008 The technology of MRI—the next 10 years? *Br. J. Radiol.* **81** 601–17

Britvitch I, Johnson I, Renker D, Stoykov A and Lorenz E 2007 Characterisation of Geiger-mode avalanche photodiodes for medical imaging applications *Nucl. Instrum. Methods Phys. Res.* A **571** 308–11

Buchanan M, Marsden P K, Mielke C H and Garlick P B 1996 A system to obtain radiotracer uptake data simultaneously with NMR spectra in a high field magnet *IEEE Trans. Nucl. Sci.* **43** 2044–8

Buonincontri G, Sawiak S J, Methner C, Krieg T, Hawkes R C and Carpenter T A 2013 PET/MRI in the infarcted mouse heart with the Cambridge split magnet *Nucl. Instrum. Methods Phys. Res.* A **702** 47–9

Burdette D *et al* 2006 *IEEE Nuclear Science Symp. Conf. Record* (*San Diego*, *29 October–4 November 2006*) vol 4 pp 2417–20

Cai L and Meng L J 2012 Hybrid pixel-waveform CdTe/CZT detector for use in an ultrahigh resolution MRI compatible SPECT system *Nucl. Instrum. Methods Phys. Res.* A **702** 101–3

Carpenter T A and Williams E J 1999 MRI—from basic knowledge to advanced strategies: hardware *Eur. Radiol.* **9** 1015–9

Catana C, van der Kouwe A, Benner T, Michel C J, Hamm M, Fenchel M, Fischl B, Rosen B, Schmand M and Sorensen A G 2010 Toward implementing an MRI-based PET attenuation-correction method for neurologic studies on the MR-PET brain prototype *J. Nucl. Med.* **51** 1431–8

Catana C, Wu Y B, Judenhofer M S, Qi J Y, Pichler B J and Cherry S R 2006 Simultaneous acquisition of multislice PET and MR images: initial results with a MR-compatible PET scanner *J. Nucl. Med.* **47** 1968–76

Cherry S R 2009 Multimodality imaging: beyond PET/CT and SPECT/CT *Semin. Nucl. Med.* **39** 348–53

Cherry S R *et al* 1997 MicroPET: a high resolution PET scanner for imaging small animals *IEEE Trans. Nucl. Sci.* **44** 1161–6

Cherry S R, Louie A Y and Jacobs R E 2008 The integration of positron emission tomography with magnetic resonance imaging *Proc. IEEE* pp 416–38

Cho Z H, Son Y D, Kim H K, Kim K N, Oh S H, Han J Y, Hong I K and Kim Y B 2007 A hybrid PET-MRI: an integrated molecular-genetic imaging system with HRRT-PET and 7.0 T MRI *Int. J. Imag. Syst. Technol.* **17** 252–65

Christensen N L, Hammer B E, Heil B G and Fetterly K 1995 Positron emission tomography within a magnetic field using photomultiplier tubes and lightguides *Phys. Med. Biol.* **40** 691–7

Defrise M, Rezaei A and Nuyts J 2012 Time-of-flight PET data determine the atttenuation sinogram up to a constant *Phys. Med. Biol.* **57** 885–99

Degenhardt C, Prescher G, Frach T, Thon A, De Gruyter R, Schmitz A and Ballizany R 2009 *IEEE Nuclear Science Symp. Conf. Record* (*Orlando, Florida, USA*, *24 October–1 November 2009*) pp 2383–6

Delso G, Furst S, Jakoby B, Ladebeck R, Ganter C, Nekolla S G, Schwaiger M and Ziegler S I 2011 Performance measurements of the siemens mMR integrated whole-body PET/MR scanner *J. Nucl. Med.* **52** 1914–22

Delso G, Martinez-Moeller A, Bundschuh R A, Ladebeck R, Candidus Y, Faul D and Ziegler S I 2010 Evaluation of the attenuation properties of MR equipment for its use in a whole-body PET/MR scanner *Phys. Med. Biol.* **55** 4361–74

Delso G and Ziegler S 2009 PET/MRI system design *Eur. J. Nucl. Med. Mol. Imag.* **36** 86–92

Disselhorst J a, Bezrukov I, Kolb A, Parl C and Pichler B J 2014 Principles of PET/MR imaging *J. Nucl. Med.* **55** 2S–10S

Dixon W T 1984 Simple proton spectroscopic imaging *Radiology* **153** 189–94

Eiber M *et al* 2011 Value of a Dixon-based MR/PET attenuation correction sequence for the localization and evaluation of PET-positive lesions *Eur. J. Nucl. Med. Mol. Imag.* **38** 1691–701

España S, Fraile L, Herraiz J, Udías J, Desco M and Vaquero J 2010 Performance evaluation of SiPM photodetectors for PET imaging in the presence of magnetic fields *Nucl. Instrum. Methods Phys. Res.* A **613** 308–16

Ford N L, Thornton M M and Holdsworth D W 2003 Fundamental image quality limits for microcomputed tomography in small animals *Med. Phys.* **30** 2869–77

Gaertner F C, Furst S and Schwaiger M 2013 PET/MR: a paradigm shift *Cancer Imag.* **13** 36–52

Graziose R, Aykac M, Casey M E, Givens G and Schmand M 2005 APD performance in light sharing PET applications *IEEE Trans. Nucl. Sci.* **52** 1413–6

Griesmer J J, Futey J, Ojha N and Morich M 2010 Whole-body PET-MR imaging system initial calibration results *2010 IEEE Nuclear Science Symp. and Medical Imaging Conf.* (*Knoxville, TN, USA*, *30 October–6 November 2010*)

Hammer B 1990 NMR-PET scanner apparatus *Magn. Reson. Imag.* **9** 4

Hammer B E, Christensen N L and Heil B G 1994 Use of a magnetic field to increase the spatial resolution of positron emission tomography *Med. phys.* **21** 1917–20

Harkness L *et al* 2011 An investigation of the performance of a coaxial HPGe detector operating in a magnetic resonance imaging field *Nucl. Instrum. Methods Phys. Res.* A **638** 67–73

Hawkes R C, Fryer T D, Lucas A J, Siegel S B, Ansorge R E, Clark J C and Carpenter T A 2008 *IEEE Nuclear Science Symp. Conf. Record* (*Dresden, Germany*, *19–25 October 2008*) pp 3673–8

Hofmann M, Bezrukov I, Mantlik F, Aschoff P, Steinke F, Beyer T, Pichler B J and Schölkopf B 2011 MRI-based attenuation correction for whole-body PET/MRI: quantitative evaluation of segmentation- and atlas-based methods *J. Nucl. Med.* **52** 1392–9 (PMID: 21828115)

Hofmann M, Steinke F, Scheel V, Charpiat G, Farquhar J, Aschoff P, Brady M, Schölkopf B and Pichler B J 2008 MRI-based attenuation correction for PET/MRI: a novel approach combining pattern recognition and atlas registration *J. Nuc. Med.* **49** 1875–83 (PMID: 18927326)

Hong K J *et al* 2013 A prototype MR insertable brain PET using tileable GAPD arrays *Med. Phys.* **40** 042503

Hong S J, Kang H G, Ko G B, Song I C, Rhee J T and Lee J S 2012 SiPM-PET with a short optical fiber bundle for simultaneous PET-MR imaging *Phys. Med. Biol.* **57** 3869–83

Jadvar H and Colletti P M 2014 Competitive advantage of PET/MRI *Eur. J. Radiol.* **83** 84–94

Judenhofer M S, Catana C, Swann B K, Siegel S B, Jung W I, Nutt R E, Cherry S R, Claussen C D and Pichler B J 2007 PET/MR images acquired with a compact MR-compatible PET detector in a 7-T magnet *Radiology* **244** 807–14

Judenhofer M S *et al* 2008 Simultaneous PET-MRI: a new approach for functional and morphological imaging *Nat. Med.* **14** 459–65

Kalemis A, Delattre B M A and Heinzer S 2013 Sequential whole-body PET/MR scanner: concept, clinical use, and optimisation after two years in the clinic. The manufacturer's perspective *Magn. Reson. Mater. Phys. Biol. Med.* **26** 5–23

Kang J, Choi Y, Hong K J, Hu W, Jung J H, Huh Y and Kim B T 2011 A small animal PET based on GAPDs and charge signal transmission approach for hybrid PET-MR imaging *J. Instrum.* **6** P08012

Karp J S, Surti S, Daube-Witherspoon M E and Muehllehner G 2008 Benefit of time-of-flight in PET: experimental and clinical results *J. Nucl. Med.* **49** 462–70

Keereman V, Fierens Y, Broux T, De Deene Y, Lonneux M and Vandenberghe S 2010 MRI-based attenuation correction for PET/MRI using ultrashort echo time sequences *J. Nucl. Med.* **51** 812–8

Keereman V, Fierens Y, Vanhove C, Lahoutte T and Vandenberghe S 2012 Magnetic resonance-based attenuation correction for micro-single-photon emission computed tomography *Mol. Imag.* **11** 155–65

Keereman V, Mollet P, Berker Y, Schulz V and Vandenberghe S 2013 Challenges and current methods for attenuation correction in PET/MR *Magn. Reson. Mater. Phys. Biol. Med.* **26** 81–98

Keereman V, Van Holen R, Mollet P and Vandenberghe S 2011 The effect of errors in segmented attenuation maps on PET quantification *Med. Phys.* **38** 6010–9

Keereman V, Vandenberghe S, De Deene Y, Luypaert R, Broux T and Lemahieu I 2008 *IEEE Nuclear Science Symp. Conf. Record* (*Dresden, Germany*, *19–25 October 2008*) pp 4656–61

Kim C, Peterson W T, Kidane T and Levin C 2014 Compensation for thermally-induced loads on PET detectors from MR stimulus in simultaneous PET/MR imaging (Abstracts of ISMRM 2014)

Kinahan P E, Hasegawa B H and Beyer T 2003 X-ray-based attenuation correction for positron emission tomography/computed tomography scanners *Semin. Nucl. Med.* **33** 166–79

Kinahan P E, Townsend D W, Beyer T and Sashin D 1998 Attenuation correction for a combined 3D PET/CT scanner *Med. Phys.* **25** 2046–53

Kolb A *et al* 2012 Technical performance evaluation of a human brain PET/MRI system *Eur. Radiol.* **22** 1776–88

Kops E R and Herzog H 2013 Errors in MR-based attenuation correction for brain imaging with PET/MR scanners *Nucl. Instrum. Methods Phys. Res.* A **702** 104–7

Kraus R, Delso G and Ziegler S I 2012 Simulation study of tissue-specific positron range correction for the new biograph mMR whole-body PET/MR system *IEEE Trans. Nucl. Sci.* **59** 1900–9

Laforest R, Longford D, Siegel S B, Newport D F and Yap J 2007 Performance evaluation of the microPET®—FOCUS-F120 *IEEE Trans. Nucl. Sci.* **54** 42–9

Levin C, Floris J, Deller T, Maramraju S H and Lagaru A 2014 Performance of a high sensitivity time-of-flight PET ring operating simultaneously within a 3T MR system *Proc. of 3rd conf. on PET-MR and SPECT-MR* (*Kos Island, Greece*, *19–21 May 2014*)

Levin C, Glover G, Deller T, McDaniel D, Peterson W and Maramraju S H 2013 Prototype time-of-flight PET ring integrated with a 3T MRI system for simultaneous whole-body PET/MR imaging *Soc. Nucl. Med. Ann. Meeting Abstracts* **54** 148

Lewellen T K 2008 Recent developments in PET detector technology *Phys. Med. Biol.* **53** R287

Liu C, Pierce L A, Alessio A M and Kinahan P E 2009 The impact of respiratory motion on tumor quantification and delineation in static PET/CT imaging *Phys. Med. Biol.* **54** 7345–62

Mackewn J E *et al* 2010 Performance evaluation of an MRI-compatible pre-clinical PET system using long optical fibers *IEEE Trans. Nucl. Sci.* **57** 1052–62

Maramraju S H *et al* 2011 Small animal simultaneous PET/MRI: initial experiences in a 9.4 T microMRI *Phys. Med. Biol.* **56** 2459–80

Martinez-Möller A and Nekolla S G 2012 Attenuation correction for PET/MR: problems, novel approaches and practical solutions *Z. Med. Phys.* **22** 299–310

Martinez-Möller A, Souvatzoglou M, Delso G, Bundschuh R A, Chefd'hotel C, Ziegler S I, Navab N, Schwaiger M and Nekolla S G 2009 Tissue classification as a potential approach for attenuation correction in whole-body PET/MRI: evaluation with PET/CT data *J. Nucl. Med.* **50** 520–6

Mollet P, Keereman V, Bini J, Izquierdo-Garcia D, Fayad Z A and Vandenberghe S 2014 Improvement of attenuation correction in time-of-flight PET/MR imaging with a positron-emitting source *J. Nucl. Med.* **55** 329–36

Mollet P, Keereman V, Clementel E and Vandenberghe S 2012 Simultaneous MR-compatible emission and transmission imaging for PET using time-of-flight information *IEEE Trans. Med. Imaging* **31** 1734–42

Muehllehner G and Karp J S 2006 Positron emission tomography *Phys. Med. Biol.* **51** R117–37

Müller-Gärtner H W, Links J M, Prince J L, Bryan R N, McVeigh E, Leal J P, Davatzikos C and Frost J J 1992 Measurement of radiotracer concentration in brain gray matter using positron emission tomography: MRI-based correction for partial volume effects *J. Cereb. Blood Flow Metab.* **12** 571–83

Nagy K, Toth M, Major P, Patay G, Egri G, Haggkvist J, Varrone A, Farde L, Halldin C and Gulyas B 2013 Performance evaluation of the small-animal nanoScan PET/MRI system *J. Nucl. Med.* **54** 1825–32

Nuyts J, Dupont P, Stroobants S, Benninck R, Mortelmans L and Suetens P 1999 Simultaneous maximum a posteriori reconstruction of attenuation and activity distributions from emission sinograms *IEEE Trans. Med. Imaging* **18** 393–403

Ouyang J, Li Q and El Fakhri G 2013 Magnetic resonance-based motion correction for positron emission tomography imaging *Semin. Nucl. Med.* **43** 60–7

Panin V Y, Aykac M and Casey M E 2013 Simultaneous reconstruction of emission activity and attenuation coefficient distribution from TOF data, acquired with external transmission source *Phys. Med. Biol.* **58** 3649–69

Peng B H and Levin C S 2010 Recent development in PET instrumentation *Curr. Pharm. Biotechnol.* **11** 555–71

Pichler B J, Judenhofer M S, Catana C, Walton J H, Kneilling M, Nutt R E, Siegel S B, Claussen C D and Cherry S R 2006 Performance test of an LSO-APD detector in a 7 T MRI scanner for simultaneous PET/MRI *J. Nucl. Med.* **47** 639–47

Pichler B J, Kolb A, Nägele T and Schlemmer H P 2010 PET/MRI: paving the way for the next generation of clinical multimodality imaging applications *J. Nucl. Med.* **51** 333–6

Pichler B, Lorenz E, Mirzoyan R, Pimpl W, Roder F, Schwaiger M and Ziegler S 1997 Performance test of a LSO-APD PET module in a 9.4 T magnet *IEEE Nuclear Science Symp. Conf. Record* (*Albuquerque, New Mexico*, *9–15 November 1997*) vol 2

Polycarpou I, Tsoumpas C, King A P and Marsden P K 2014 Impact of respiratory motion correction and spatial resolution on lesion detection in PET: a simulation study based on real MR dynamic data *Phys. Med. Biol.* **59** 697–713

Polycarpou I, Tsoumpas C and Marsden P K 2012 Analysis and comparison of two methods for motion correction in PET imaging *Med. Phys.* **39** 6474

Poole M, Bowtell R, Green D, Pittard S, Lucas A, Hawkes R and Carpenter A 2009 Split gradient coils for simultaneous PET-MRI *Magn. Reson. Med.* **62** 1106–11

Rahmim A, Rousset O and Zaidi H 2007 Strategies for motion tracking and correction in PET *PET Clin.* **2** 251–66

Raylman R R, Majewski S, Lemieux S K, Velan S S, Kross B, Popov V, Smith M F, Weisenberger A G, Zorn C and Marano G D 2006 Simultaneous MRI and PET imaging of a rat brain *Phys. Med. Biol.* **51** 6371–9

Renker D 2007 New trends on photodetectors *Nucl. Instrum. Methods Phys. Res.* A **571** 1–6

Rezaei A, Defrise M, Bal G, Michel C, Conti M, Watson C and Nuyts J 2012 Simultaneous reconstruction of activity and attenuation in time-of-flight PET *IEEE Trans. Med. Imaging* **31** 2224–33

Roncali E and Cherry S R 2011 Application of silicon photomultipliers to positron emission tomography *Ann. Biomed. Eng.* **39** 1358–77

Rousset O, Rahmim A, Alavi A and Zaidi H 2007 Partial volume correction strategies in PET *PET Clin.* **2** 235–49

Salomon A, Goedicke A, Schweizer B, Aach T and Schulz V 2011 Simultaneous reconstruction of activity and attenuation for PET/MR *IEEE Trans. Med. Imaging* **30** 804–13

Samarin A, Kuhn F P, Crook D W, Wiesinger F, Wollenweber S D, von Schulthess G K, Hodler J and Schmid D T 2011 Image registration accuracy of a sequential, tri-modality PET/CT plus MR imaging setup using dedicated patient transporter systems *Eur. J. Nucl. Med. Mol. Imag.* **38** S220

Schaart D R, Seifert S, Vinke R, van Dam H T, Dendooven P, Löhner H and Beekman F J 2010 LaBr(3):Ce and SiPMs for time-of-flight PET: achieving 100 ps coincidence resolving time *Phys. Med. Biol.* **55** N179–89

Schenck J F 1996 The role of magnetic susceptibility in magnetic resonance imaging: MRI magnetic compatibility of the first and second kinds *Med. Phys.* **23** 815–50

Schenck J F, Jolesz F A, Roemer P B, Cline H E, Lorensen W E, Kikinis R, Silverman S G, Hardy C J, Barber W D and Laskaris E T 1995 Superconducting open-configuration MR imaging system for image-guided therapy *Radiology* **195** 805–14

Schulz V *et al* 2009 A preclinical PET/MR insert for a human 3 T MR scanner *IEEE Nuclear Science Symp. and Medical Imaging Conf.* (*Orlando, Florida, USA*, *24 October–1 November 2009*)

Shah N J *et al* 2013 Advances in hybrid MR-PET at 3 T and 9.4 T in humans *Nucl. Instrum. Methods Phys. Res.* A **702** 16–21

Shao Y, Cherry S R, Farahani K, Slates R, Silverman R W, Meadors K, Bowery A, Siegel S, Marsden P K and Garlick P B 1997 Development of a PET detector system compatible with MRI/NMR systems *IEEE Trans. Nucl. Sci.* **44** 1167–71

Shao Y, Cherry S R, Siegel S, Silverman R W and Marsden P K 1996 Feasibility study of high resolution PET detectors for imaging in high magnetic field environments *J. Nucl. Med.* **37** 330

Shao Y 1997 Simultaneous PET and MR imaging *Phys. Med. Biol.* **42** 1965–70

Somer E J, Benatar N A, O'Doherty M J, Smith M A and Marsden P K 2007 Use of the CT component of PET-CT to improve PET-MR registration: demonstration in soft-tissue sarcoma *Phys. Med. Biol.* **52** 6991–7006

Stanisz G J, Odrobina E E, Pun J, Escaravage M, Graham S J, Bronskill M J and Henkelman R M 2005 T1, T2 relaxation and magnetization transfer in tissue at 3T *Magn. Reson. Med.* **54** 507–12

Steinberg J, Jia G, Sammet S, Zhang J, Hall N and Knopp M V 2010 Three-region MRI-based whole-body attenuation correction for automated PET reconstruction **37** 227–35

Strul D, Cash D, Keevil S F, Halsted P, Williams S C R and Marsden P K 2003 Gamma shielding materials for MR-compatible PET *IEEE Trans. Nucl. Sci.* **50** 60–9

Turkington T G 2000 Attenuation correction in hybrid positron emission tomography *Semin. Nucl. Med.* **30** 255–67

Veit-Haibach P, Kuhn F P, Wiesinger F, Delso G and von Schulthess G 2013 PET-MR imaging using a tri-modality PET/CT-MR system with a dedicated shuttle in clinical routine *Magn. Reson. Mater. Phys. Biol. Med.* **26** 25–35

Vunckx K, Atre A, Baete K, Reilhac A, Deroose C M and Van Laere K and Nuyts J 2012 Evaluation of three MRI-based anatomical priors for quantitative PET brain imaging *IEEE Trans. Med. Imaging* **31** 599–612

Wagadarikar I and Dolinsky M 2014 Sensitivity improvement of time-of-flight (ToF) PET detector through recovery of compton scattered annihilation photons *IEEE Trans. Nucl. Sci.* **61** 121–5

Watson C C, Eriksson L and Kolb A 2013 Physics and applications of positron beams in an integrated PET/MR *Phys. Med. Biol.* **58** L1–2

Wehner J, Weissler B, Dueppenbecker P, Gebhardt P, Schug D, Ruetten W, Kiessling F and Schulz V 2014 PET/MRI insert using digital SiPMs: investigation of MR-compatibility *Nucl. Instrum. Methods Phys. Res.* A **734** 116–21

Wehrl H F, Judenhofer M S, Thielscher A, Martirosian P, Schick F and Pichler B J 201 Assessment of MR compatibility of a PET insert developed for simultaneous multiparametric PET/MR imaging on an animal system operating at 7 T *Magn. Reson. Med.* **65** 269–79

Weirich C, Brenner D, Scheins J, Besancon E, Tellmann L, Herzog H and Shah N J 2012 Analysis and correction of count rate reduction during simultaneous MR-PET measurements with the brain PET scanner *IEEE Trans. Med. Imaging* **31** 1372–80

Weissler B *et al* 2012 Design concept of world's first preclinical PET/MR insert with fully digital silicon photomultiplier technology *IEEE Nuclear Science Symp. and Medical Imaging Conf. Record* (*Anaheiim, CA, USA*, *29 October–3 November 2012*) B Yu (Danvers: IEEE) pp 2113–6

Weissler B *et al* 2014 MR compatibility aspects of a silicon photomultiplier-based PET/RF insert with integrated digitisation *Phys. Med. Biol.* **59** 5119–39

Yamamoto S, Imaizumi M, Kanai Y, Tatsumi M, Aoki M, Sugiyama E, Kawakami M, Shimosegawa E and Hatazawa J 2010 Design and performance from an integrated PET/MRI system for small animals *Ann. Nucl. Med.* **24** 89–98

Yamamoto S, Kuroda K and Senda M 2003 Scintillator selection for MR-compatible gamma detectors *IEEE Trans. Nucl. Sci.* **50** 1683–5

Zaidi H, Ojha N, Morich M, Griesmer J, Hu Z, Maniawski P, Ratib O, Izquierdo-Garcia D, Fayad Z A and Shao L 2011 Design and performance evaluation of a whole-body ingenuity TF PET-MRI system *Phys. Med. Biol.* **56** 3091–106